\documentclass[twocolumn,twocolappendix,trackchanges]{aastex63}

\newcommand{\kms}{\mathrm{km\,s}^{-1}}

\usepackage{amsmath}	% Advanced maths commands
\usepackage{amssymb}	% Extra maths symbols
\usepackage{multirow}

\received{2021 November 15}
\revised{2021 December}
\accepted{2021 December 6}
\submitjournal{ApJS}

\shorttitle{SB2s from LAMOST MRS}
\shortauthors{Zhang et al.}
\graphicspath{{./}{figures/}}
\begin{document}

\title{The Spectroscopic Binaries from LAMOST Medium-Resolution Survey (MRS). I.\\
Searching for Double-lined Spectroscopic Binaries (SB2s) with Convolutional Neural Network}

\correspondingauthor{Bo Zhang (LAMOST Fellow)}
\email{bozhang@bnu.edu.cn}

%\correspondingauthor{Bo Zhang}
%\email{bozhang@bnu.edu.cn}

\author[0000-0002-6434-7201]{Bo Zhang}
\affiliation{Department of Astronomy, Beijing Normal University, Beijing 100875, People's Republic of China}
% \collaboration{1}{(LAMOST Fellow)}
% \affiliation{Key Laboratory of Optical Astronomy, National Astronomical Observatories, Chinese Academy of Sciences, Beijing 100101,  People's Republic of China}
% \affiliation{University of Chinese Academy of Sciences, Beijing 100049, People's Republic of China}

\author[0000-0003-3433-8416]{Ying-Jie Jing}
\affiliation{Key Laboratory for Computational Astrophysics, National Astronomical Observatories, Chinese Academy of Sciences, Beijing 100101,  People's Republic of China}
%\affiliation{School of Astronomy and Space Science, University of Chinese Academy of Sciences, Beijing, 100049, China}

\author[0000-0002-6039-8212]{Fan Yang}
\affiliation{Department of Astronomy, Beijing Normal University, Beijing 100875, People's Republic of China}
%\affiliation{Key Laboratory of Optical Astronomy, National Astronomical Observatories, Chinese Academy of Sciences, Beijing 100101,  People's Republic of China}
%\affiliation{School of Astronomy and Space Science, University of Chinese Academy of Sciences, Beijing, 100049, China}

\author[0000-0003-1831-2005]{Jun-Chen Wan}
%\affiliation{Alibaba group, Beijing 100102,  People's Republic of China}
\affiliation{Kuaishou Technology, Beijing 100085,  People's Republic of China}

\author[0000-0003-4829-6245]{Xin Ji}
\affiliation{Key Laboratory of Optical Astronomy, National Astronomical Observatories, Chinese Academy of Sciences, Beijing 100101,  People's Republic of China}
\affiliation{School of Astronomy and Space Science, University of Chinese Academy of Sciences, Beijing, 100049, People's Republic of China}

%\author[0000-0001-7790-5498]{Meng-Ting Liu}
%\affiliation{Department of Astronomy, Beijing Normal University, Beijing 100875, People's Republic of China}

\author[0000-0001-8241-1740]{Jian-Ning Fu}
\affiliation{Department of Astronomy, Beijing Normal University, Beijing 100875, People's Republic of China}

\author[0000-0002-1802-6917]{Chao Liu}
\affil{Key Laboratory of Space Astronomy and Technology, National Astronomical Observatories, Chinese Academy of Sciences, Beijing 100101, People's Republic of China}

\author[0000-0002-5164-3773]{Xiao-Bin Zhang}
\affiliation{Key Laboratory of Optical Astronomy, National Astronomical Observatories, Chinese Academy of Sciences, Beijing 100101,  People's Republic of China}

\author[0000-0003-0615-4695]{Feng Luo}
\affiliation{Key Laboratory of Optical Astronomy, National Astronomical Observatories, Chinese Academy of Sciences, Beijing 100101,  People's Republic of China}
\affiliation{School of Astronomy and Space Science, University of Chinese Academy of Sciences, Beijing, 100049, People's Republic of China}

\author[0000-0003-3347-7596]{Hao Tian}
\affil{Key Laboratory of Space Astronomy and Technology, National Astronomical Observatories, Chinese Academy of Sciences, Beijing 100101, People's Republic of China}

\author[0000-0002-4391-2822]{Yu-Tao Zhou}
\affil{Department of Astronomy, School of Physics, Peking University, Beijing 100871, People’s Republic of China}
\affil{Kavli Institute for Astronomy and Astrophysics, Peking University, Beijing 100871, People’s Republic of China}

\author[0000-0002-6868-6809]{Jia-Xin Wang}
\affil{Department of Astronomy, Beijing Normal University, Beijing 100875, People's Republic of China}

\author[0000-0001-9989-9834]{Yan-Jun Guo}
\affiliation{Yunnan observatories, Chinese Academy of Sciences, Kunming, 650011, People's Republic of China}
\affiliation{Key Laboratory for Structure and Evolution of Celestial Objects, Chinese Academy of Sciences, Kunming 650216, People's Republic of China}
\affiliation{School of Astronomy and Space Science, University of Chinese Academy of Sciences, Beijing, 100049, People's Republic of China}

\author[0000-0002-7660-9803]{Weikai Zong}
\affiliation{Department of Astronomy, Beijing Normal University, Beijing 100875, People's Republic of China}

\author[0000-0003-4829-6245]{Jian-Ping Xiong}
\affiliation{Key Laboratory of Optical Astronomy, National Astronomical Observatories, Chinese Academy of Sciences, Beijing 100101,  People's Republic of China}
\affiliation{School of Astronomy and Space Science, University of Chinese Academy of Sciences, Beijing, 100049, People's Republic of China}

\author[0000-0002-2577-1990]{Jiao Li}
\affiliation{Key Laboratory of Optical Astronomy, National Astronomical Observatories, Chinese Academy of Sciences, Beijing 100101,  People's Republic of China}

\begin{abstract}
We developed a convolutional neural network (CNN) model to distinguish the double-lined spectroscopic binaries (SB2s) from others based on single exposure medium-resolution spectra ($R\sim 7,500$).
The training set consists of a large set of mock spectra of single stars and binaries synthesized based on the MIST stellar evolutionary model and ATLAS9 atmospheric model.
Our model reaches a novel theoretic false positive rate by adding a proper penalty on the negative sample (e.g., 0.12\% and 0.16\% for the blue/red arm when the penalty parameter $\Lambda=16$).
Tests show that the performance is as expected and favors FGK-type Main-sequence binaries with high mass ratio ($q \geq 0.7$) and large radial velocity separation ($\Delta v \geq 50\,\kms$).
Although the real false positive rate can not be estimated reliably, validating on eclipsing binaries identified from Kepler light curves indicates that our model predicts low binary probabilities at eclipsing phases (0, 0.5, and 1.0) as expected. 
The color-magnitude diagram also helps illustrate its feasibility and capability of identifying FGK MS binaries from spectra.
We conclude that this model is reasonably reliable and can provide an automatic approach to identify SB2s with period $\lesssim 10$ days.
This work yields a catalog of binary probabilities for over 5 million spectra of 1 million sources from the LAMOST medium-resolution survey (MRS), and a catalog of 2198 SB2 candidates whose physical properties will be analyzed in our following-up paper. 
Data products are made publicly available at the journal as well as our Github website.

\end{abstract}



\section{Introduction} \label{sec:intro}

Binary systems, as a cornerstone of modern astrophysics, are as common as single stars across the color-magnitude diagram \citep{1969JRASC..63..275H, 1976ApJS...30..273A, 1991A&A...248..485D, 2010ApJS..190....1R, 2013ARA&A..51..269D}.
Despite many problems including the formation of sources of gravitational waves \citep{2001MNRAS.324..797S} and black hole binaries \citep{2019Natur.575..618L} rely on the understanding of binaries, binaries themselves also present many unresolved problems, such as their formation, evolution, and interactions \citep{2001icbs.book.....H,  2020RAA....20..161H}.
The binary fraction of field stars and its correlation with chemical abundances are extensively studied in recent years \citep{2014ApJ...788L..37G, 2017MNRAS.469L..68G, 2015ApJ...799..135Y, 2018RAA....18...52T, 2019ApJ...875...61M, 2020MNRAS.499.1607M, 2021arXiv210904031N}.
The statistical distributions of masses orbital periods, mass ratios and eccentricities are not well known yet \citep{1991A&A...248..485D,2003A&A...397..159H, 2017ApJS..230...15M}.

Binaries can be identified and characterized by astrometric, photometric, and spectroscopic observations, while different observation techniques favor different kinds of binaries \citep{2017ApJS..230...15M}.
Beyond the solar neighborhood ($d>5\,\mathrm{pc}$), most binaries remain unresolved with earth-based observations due to distances \citep{2006epbm.book.....E}.
Therefore, spectroscopic observations remain a valuable method that can characterize distant binaries.
Large spectroscopic surveys provide good opportunities to mine spectroscopic binaries (SBs).
Conventionally, SBs can be divided into two categories, i.e., single-lined spectroscopic binaries (SB1s) and double-lined spectroscopic binaries (SB2s) depending on whether the secondary is visible in spectra.
As a consequence, SB1s are usually identified based on variability of radial velocities \citep{2010AJ....140..184M, 2020ApJS..249...31Y} while SB2s are identified with double peaks in cross-correlation function \citep[CCF,][]{2011AJ....141..200M,2017A&A...608A..95M, 2021ApJS..256...31L, 2021AJ....162..184K}.
Efforts are also made to utilize machine learning methods to identify SB2s \citep[e.g., t-SNE,][]{2017ApJS..228...24T,2020A&A...638A.145T} and unresolved SBs \citep[data-driven spectral model,][]{2018MNRAS.473.5043E, 2018MNRAS.476..528E}.

Most of the existing large ground-based spectroscopic surveys, such as LAMOST low-resolution survey \citep[LRS, $R\sim1\,800$,][]{2012RAA....12.1197C,2012RAA....12..735D}, APOGEE \citep{2017AJ....154...94M}, GALAH \citep{2015MNRAS.449.2604D}, and RAVE \citep{2006AJ....132.1645S} typically observe an object only a few times, allowing discovery and characterization of binaries but makes orbital solutions very difficult.
Due to short time baseline and limited spectral resolution, most discovered spectroscopic binaries are close binaries \citep[$P<$ a few years and $a<$ a few AUs, e.g.,][]{2020ApJ...895....2P}.

The LAMOST medium-resolution survey \citep[MRS, $R\sim7\,500$,][]{2020arXiv200507210L}, started from Sep. 2017, is one of the pioneer time-domain spectroscopic surveys.
The blue and red arms cover Mg $b$ triplet ($4950-5350~\mathrm{\AA}$)  and H$\alpha$ ($6300-6800~\mathrm{\AA}$), respectively.
The targeting is based on Gaia \citep{2016A&A...595A...1G}, and the limiting magnitude is around $G\sim15$ mag.
The radial velocities can be determined to a precision of $\sim 0.85\, \kms$ for high S/N spectra \citep{2021ApJS..256...14Z}.
The LAMOST MRS DR8 already contains $\sim 5$ million single exposure spectra (S/N$>5$) for 1 million sources which are obtained from Sep. 2017 to Jun. 2020.
The survey has extensively observed the \textit{Kepler/K2} \citep{2010Sci...327..977B,2014PASP..126..398H} and \textit{TESS} fields \citep{2015JATIS...1a4003R}.
In particular, the LAMOST-MRS-\textit{Kepler/K2} project will take $\sim80$ exposures for over 50\,000 stars selected from \textit{Kepler/K2} survey in the five-year MRS period \citep{2020RAA....20..167F, 2020ApJS..251...15Z}.
Combined with high-precision photometry from \textit{Kepler/K2} and \textit{TESS}, it forms a valuable and sizable dataset for time-domain research on this scientific goal \citep[e.g.,][]{2020ApJ...905...67P, 2021PASP..133d4202P, 2021MNRAS.504.4302W, 2021MNRAS.506.6117W}.
A subproject towards 4 \textit{K2} plates, namely the LAMOST-TD survey \citep{2021arXiv210903149W}, has led to the discovery of a subgiant with an undetected 1-3 $M_\odot$ companion \citep{2021arXiv211007944Y}.
Minute-cadence photometry is obtained for LAMOST MRS fields as well to study short period variables \citep{2021MNRAS.tmp.2562L}.

Our previous paper \citep{2021ApJS..256...14Z} has presented self-consistent radial velocity (RV) measurements calibrated to \textit{Gaia} DR2 scale for LAMOST MRS DR7.
%In order to study the physical properties of SBs, we need a sample of spectroscopic binaries as our firm basis.
In this paper, we build an efficient discriminative model based on convolutional neural network (CNN) to distinguish SB2s from SB1s/single stars with LAMOST MRS DR8 single exposure spectra.
In Section \ref{sec:training_set}, we describe the synthesis of our training set.
In Section \ref{sec:model}, we show the training process of the CNN discriminative model and its performance on the synthetic test set.
In Section \ref{sec:results}, we show the results and validate with stars in solar neighborhood ($\varpi>3$) and \textit{Kepler} eclipsing binaries as well as the catalog of SB2 candidates.
A brief discussion is presented in Section \ref{sec:discussion}.
We summarize this work in Section \ref{sec:conclusions}.

\begin{deluxetable*}{cccc}
\tablenum{1}
\tablecaption{A summary of the free parameters ($\boldsymbol{\theta}$ and $\mathrm{S/N}$).\label{tab:params}}
\tablewidth{0pt}
\tablehead{
\colhead{Quantity} & \colhead{Distribution}  & \colhead{Unit} & \colhead{Notes}}
% \decimalcolnumbers
\startdata
\multirow{3}{*}{$m_1$} & \multirow{3}{*}{$p(m_1) \propto \begin{cases} 1 & 0.1<m_1/M_\odot < 1  \\ m_1^{-2.35} & 1<m_1/M_\odot < 2\end{cases}$} & \multirow{3}{*}{$M_\odot$} & \multirow{3}{*}{to balance spectral types}\\
\multirow{3}{*}{} & \multirow{3}{*}{} & \multirow{3}{*}{} & \multirow{3}{*}{}\\
\multirow{3}{*}{} & \multirow{3}{*}{} & \multirow{3}{*}{} & \multirow{3}{*}{}\\
$q$ &  $\boldsymbol{\mathcal{U}}(0.01,\,1)$&   & $ q \equiv m_2 / m_1 $\\
$\mathrm{[Fe/H]}$ & $\boldsymbol{\mathcal{U}}(-2.0,\,0.5)$ & dex & \\
$\mathrm{[\alpha/Fe]}$ & $\boldsymbol{\mathcal{U}}(-0.5,\,0.7)$ & dex & \\
EEP$_1$ & $\boldsymbol{\mathcal{U}}(202,\,605)$  &  & from ZAMS to RGB tip\\
$(v_\mathrm{eq}\sin{i})_{1/2}$ &  $\boldsymbol{\mathcal{U}}(0,\,500)$ & km s$^{-1}$ & \\
%$(v_\mathrm{eq}\sin{i})_2$ &  $\boldsymbol{\mathcal{U}}(0, 500)$ & km s$^{-1}$ & \\
$v_1$ & $\boldsymbol{\mathcal{N}}(0,\,5^2)$ & km s$^{-1}$ & for dithering\\
$\Delta v$ & $\boldsymbol{\mathcal{U}}(-500,\,500)$ & km s$^{-1}$ & $ \Delta \equiv v_2-v_1$\\
\hline 
S/N & $\boldsymbol{\mathcal{U}}(5,\,100)$  &  & \\
% $T_\mathrm{eff}$ and $\log{g}$ & derived from MIST model  &  & \\
\enddata
\tablecomments{$\boldsymbol{\mathcal{U}}(a, b)$ represents a uniform distribution from $a$ to $b$, and $\boldsymbol{\mathcal{N}}(\mu, \sigma^2)$ represents a normal distribution with mean $\mu$ and variance $\sigma^2$.}
\end{deluxetable*}

\begin{deluxetable*}{c|ccccc}
\tablenum{2}
\tablecaption{A summary of the training set.\label{tab:trainingset}}
\tablewidth{0pt}
\tablehead{
\colhead{Sample} & Label $y$ & Weight & \colhead{$\mathrm{N_{train}}$}  & \colhead{$\mathrm{N_{validation}}$} & \colhead{$\mathrm{N_{test}}$}}
% \decimalcolnumbers
\startdata
$\{\boldsymbol{F}_\mathrm{b}(\boldsymbol{\theta})\}$ & $1$ & $1$ & 306\,000 & 34\,000 & 40\,000 \\
%$[\boldsymbol{F}_\mathrm{binary}(\boldsymbol{\theta})]_\mathrm{S/N}^\mathrm{DA}$ & $1$ & 80\,000 & 20\,000 & 10\,000 \\
$\{\boldsymbol{F}_1(\boldsymbol{\phi}_1)$\} & $0$ & $\Lambda$ & 306\,000 & 34\,000 & 40\,000 \\
%$[\boldsymbol{F}_1(\boldsymbol{\phi}_1)]_\mathrm{S/N}^\mathrm{DA}$ & $0$ & 80\,000 & 20\,000 & 10\,000 \\
%$\boldsymbol{F}^\mathrm{DA}$ & $0$ & 80\,000 & 20\,000 & 10\,000 \\
\enddata
%\tablecomments{$\boldsymbol{\mathcal{U}}(a, b)$ represents a uniform distribution from $a$ to $b$, and $\boldsymbol{\mathcal{N}}(\mu, \sigma^2)$ represents a normal distribution with mean $\mu$ and variance $\sigma^2$.}
\end{deluxetable*}

\begin{deluxetable*}{r|cccc} \label{tab:acc}
\tablenum{3}
\tablecaption{The mean TPR and FPR for models with different $\Lambda$.}
\tablewidth{0pt}
\tablehead{
\colhead{} & \multicolumn2c{Blue arm} & \multicolumn2c{Red arm} \\
\colhead{$\Lambda$} & Mean TPR & Mean FPR & Mean TPR & Mean FPR }
% \decimalcolnumbers
\startdata
$1$ & 0.679015 & 0.183200 & 0.628980 & 0.151890 \\
$2$ & 0.496245 & 0.036595 & 0.483435 & 0.041970 \\
$4$ & 0.436340 & 0.011250 & 0.413220 & 0.011075 \\
$8$ & 0.393160 & 0.004140 & 0.368120 & 0.004380 \\
$16$ & 0.359925 & 0.001275 & 0.347165 & 0.001620 \\
$32$ & 0.337955 & 0.000355 & 0.312365 & 0.000645 \\
$64$ & 0.318940 & 0.000205 & 0.302915 & 0.000280 \\
$128$ & 0.311070 & 0.000155 & 0.276590 & 0.000125 \\
$256$ & 0.286845 & 0.000055 & 0.268060 & 0.000050 \\
$512$ & 0.281780 & 0.000010 & 0.256640 & 0.000030 \\
\enddata
%\tablecomments{$\boldsymbol{\mathcal{U}}(a, b)$ represents a uniform distribution from $a$ to $b$, and $\boldsymbol{\mathcal{N}}(\mu, \sigma^2)$ represents a normal distribution with mean $\mu$ and variance $\sigma^2$.}
\end{deluxetable*}

\section{The Training Set} \label{sec:training_set}
To construct a large and homogeneous training set of labeled spectra (label=0 for single stars and label=1 for binaries) across the parameter space, we use the MIST stellar evolutionary tracks \citep{2016ApJS..222....8D, 2016ApJ...823..102C} and a grid of synthetic stellar spectra \citep{2018A&A...618A..25A} to synthesize our training set.
Assuming that the primary and secondary stars in a (primordial) binary system share the same elemental abundance and age, the vector of free parameters for a binary system is
\begin{equation}
\begin{aligned}
	\boldsymbol{\theta} = \{ m_1, & q, \mathrm{EEP}_1, \mathrm{[Fe/H]}, \mathrm{[\alpha/Fe]}, \\
	  & (v_\mathrm{eq}\sin{i})_1, (v_\mathrm{eq}\sin{i})_2, v_1, \Delta v \},
\end{aligned}
\end{equation}
where $m_1$ is the mass of the primary star, $q\equiv m_2/m_1$ is the mass ratio ($m_2 \leq m_1$), $\mathrm{EEP}_1$ is the Equivalent Evolutionary Point \citep[defined in][]{2016ApJS..222....8D} value of the primary star covering from Zero-age main sequence (ZAMS) to Red giant brach tip (RGB tip) which is used as a proxy of age, $\mathrm{[Fe/H]}$ and $\mathrm{[\alpha/Fe]}$ are the iron abundance and $\alpha$-element abundance, $(v_\mathrm{eq}\sin{i})_{1/2}$ is the projected rotational velocity of corresponding component, $v_1$ is the radial velocity of the primary star, $\Delta v \equiv v_2-v_1 $ is the radial velocity separation, and $\mathrm{S/N}$ is the signal-to-noise ratio of the synthetic spectrum.
Note that, the adopted Salpeter IMF \citep{ 1955ApJ...121..161S} is flatten to reduce the number of low-mass stars,  the adoption of $\mathrm{EEP}_1$ instead of age is to cover the $T_\mathrm{eff}-\log{g}$ parameter space evenly, the upper boundary of $(v_\mathrm{eq}\sin{i})_{1/2}$ is set to $500\,\kms$ to cover the majority of the Galactic stars \citep{ 2020ApJ...892L..26L}, and the upper boundary of $\Delta v$ is set to be higher than the maximum radial velocity separation for a solar twin binary with circular orbits (i.e., $m_1=m_2=1M_\odot$, $e=0$, and $i=90^\circ$) and extremely short period $P=10^{-3}\,\mathrm{day}$.
We assume that the observed spectrum of a binary is dominated by the features of the primary star, so that we can measure $v_1$ (the RV of primary star) from a spectrum and shift the spectrum to its rest-of-frame\footnote{This is not always true. When $q\sim 1$, it is possible that the measured RV is that of the secondary star. But this does not affect our training.}, so $v_1$ is drawn with small deviation from zero considering that RV measurements have errors. 
The S/N of synthetic spectra are drawn from a uniform distribution ranging from 5 to 100 which are typical for the LAMOST MRS data.
Table \ref{tab:params} summarizes how we draw these free parameters.

We use the MIST stellar evolutionary tracks to convert $\boldsymbol{\theta}$ to the stellar parameters of primary and secondary stars, i.e., 
\begin{equation}
	\boldsymbol{\phi}_1(\boldsymbol{\theta}) = \{ T_{\mathrm{eff},1}, \log{g}_1, \mathrm{[Fe/H]}, \mathrm{[\alpha/Fe]}, (v_\mathrm{eq} \sin{i})_1, v_1\} ,
\end{equation}
and 
\begin{equation}
	\boldsymbol{\phi}_2(\boldsymbol{\theta}) = \{ T_{\mathrm{eff},2}, \log{g}_2, \mathrm{[Fe/H]}, \mathrm{[\alpha/Fe]}, (v_\mathrm{eq} \sin{i})_2, v_2\} .
\end{equation}
Meanwhile, we also evaluate the radii ($R_{1/2}$), luminosities ($L_{1/2}$) and ages ($\tau$).
Note that, during this conversion, $\mathrm{EEP}_2$ is determined that the primary and secondary stars have the same age.
Then we interpolate the synthetic spectral library which is already degraded to $R=7\,500$ to generate spectra with $\boldsymbol{\phi_1}$ and $\boldsymbol{\phi_2}$.
The binary spectrum is, therefore, a sum of the two components' spectra,
\begin{equation}
[\boldsymbol{F}_\mathrm{binary}(\boldsymbol{\theta})]_\mathrm{S/N} = [\boldsymbol{F}_1 (\boldsymbol{\phi_1}) \times R_1^2+\boldsymbol{F}_2(\boldsymbol{\phi_2}) \times R_2^2]_\mathrm{S/N}, 
\end{equation}
where $[\cdot]_\mathrm{S/N}$ represents the operation of adding Gaussian noise to degrade the spectrum to a given $\mathrm{S/N}$.
For each binary spectrum $[\boldsymbol{F}_\mathrm{binary}(\boldsymbol{\theta})]_\mathrm{S/N}$, we also calculate the spectrum of its primary star, i.e., $[\boldsymbol{F}_1 (\boldsymbol{\phi_1})]_\mathrm{S/N}$, as our negative sample.
All the spectra are sampled to the same wavelength grid with approximately 0.1\,$\mathrm{\AA}$ step in blue arm $5\,000-5\,300\,\mathrm{\AA}$ (3\,347 pixels) and red arm $6\,350-6\,750\,\mathrm{\AA}$ (3\,509 pixels), and then normalized to continuum using an iterative fitting with smoothing spline \citep{2021ApJS..256...14Z}.

Note that our single stars are just the primary stars of the corresponding binaries, so our training set consists of a large number of single star-binary pairs.
The advantage of such a training set is that  for each binary the model knows how the spectrum of its primary star looks like.
By imposing proper penalties on negative samples, as long as the contribution from the secondary is not significant, the spectrum of the binary will be classified as a single star.
As a result, only the binaries with prominent features from secondaries, which we call SB2s, will be picked out.
Starting from 1 million initial samples, about 380\,000 single star-binary pairs survived.
The majority of the initial samples are dropped mainly due to that at least one of the two stars goes out of the range of the synthetic spectral library.
For example, if the secondary star is a dwarf star with a very low mass, its effective temperature will be below 3500K which is the limit of our synthetic library and we can not synthesize the binary spectrum.
Finally, we separate 40\,000 single star-binary pairs for testing our model, and the rest 340\,000 pairs are split via 9:1 for training and validating our models.
Table \ref{tab:trainingset} has summarized how we split the sample.

\begin{figure*}
\plotone{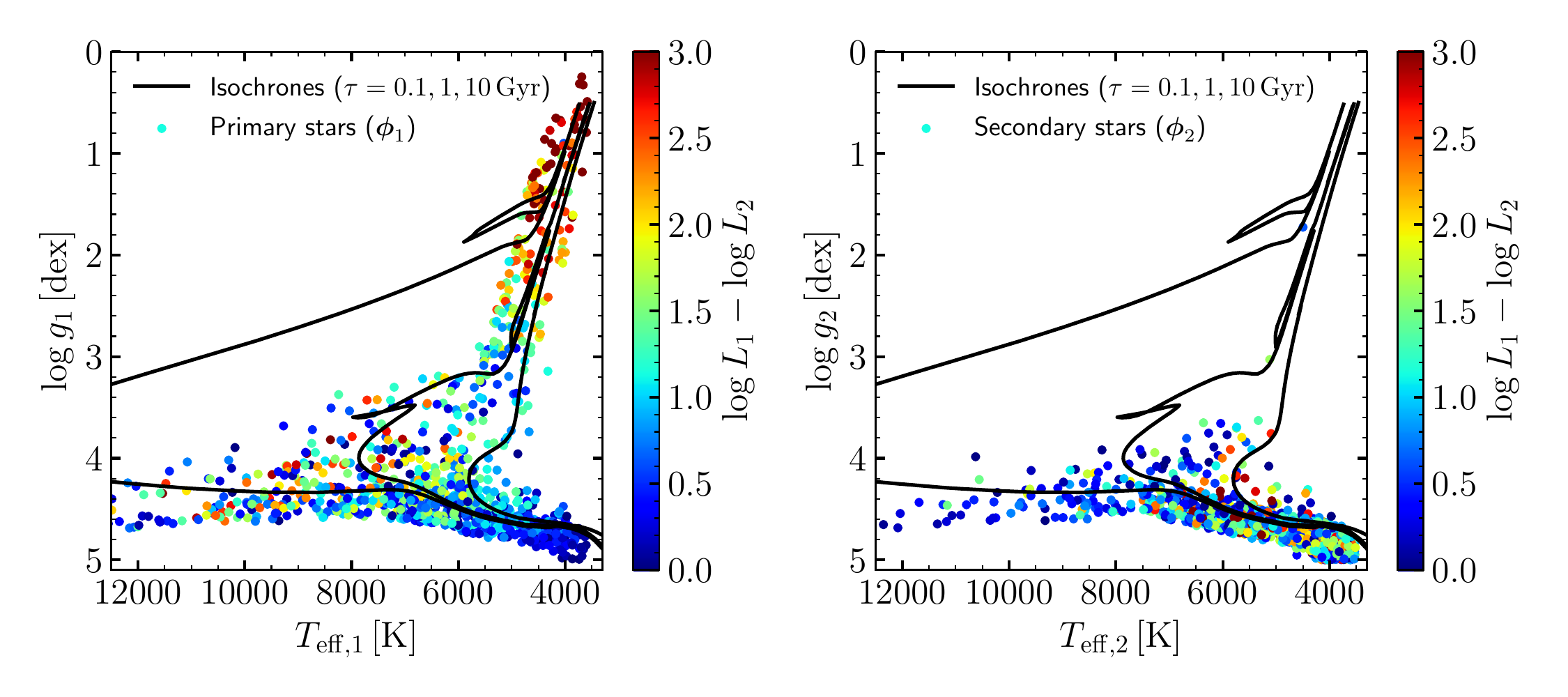}
\caption{The stellar parameters of 1000 randomly chosen mock binaries in our training set.
The left and right panels show primary stars and secondary stars, respectively.
Color codes the luminosity ratios.
Superposed are a set of isochrones with solar metallicity but different ages.
\label{fig:tefflogg}}
\end{figure*}

\begin{figure*}
\plotone{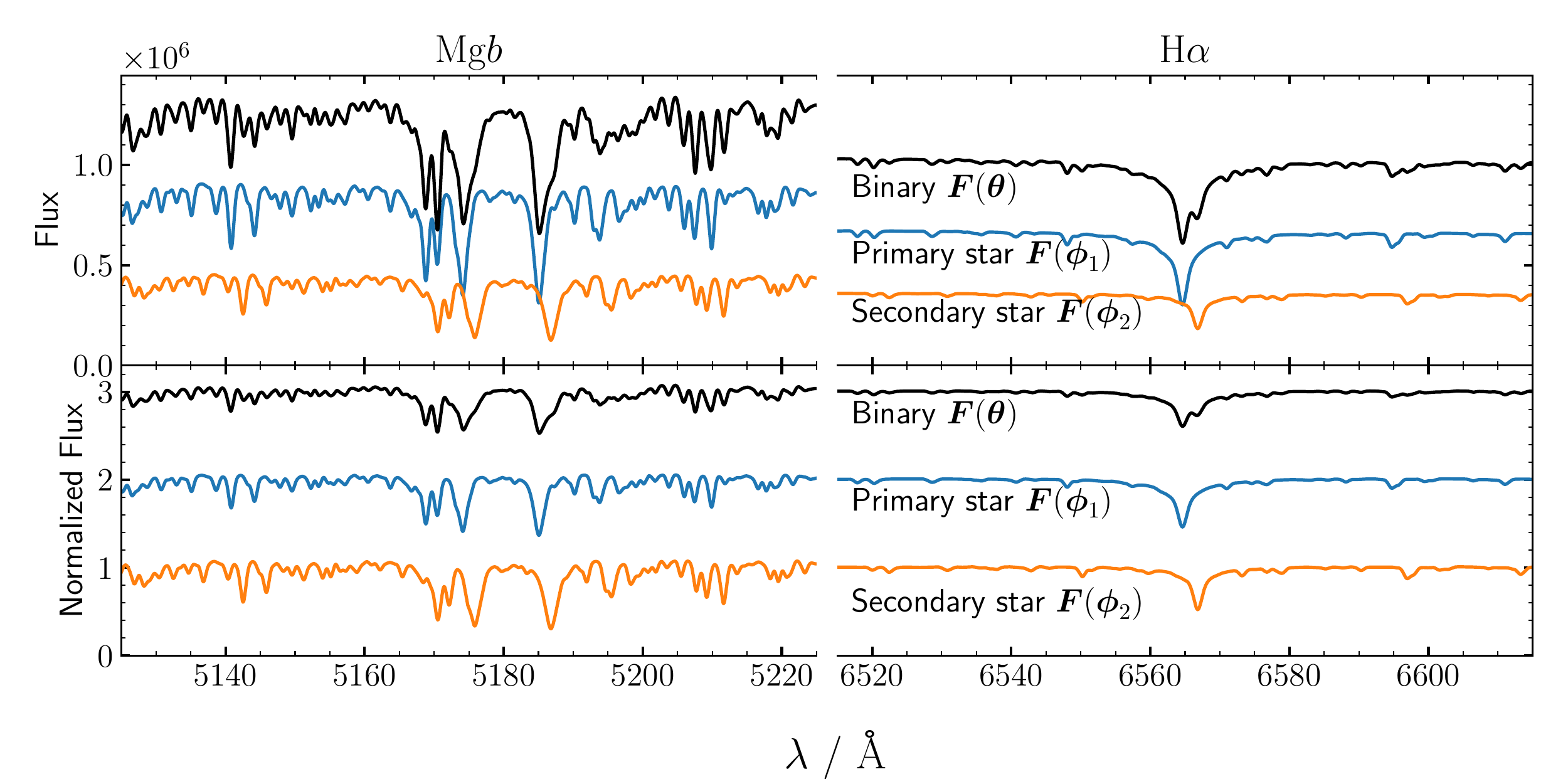}
\caption{This figure demonstrates how binary spectra are synthesized in our experiment.
The upper and lower panel shows the relatively flux and normalized flux, respectively.
The initial mass of the primary star is 1$M_\odot$ and that of the secondary is 0.9 $M_\odot$ ($q=0.9$).
Both of them have solar metallicity and 5 Gyr age.
A 100 km s$^{-1}$ radial velocity separation is applied.
Differences could be found by comparing the spectra of the primary star and the binary, which indicates that a classification is possible to some extent.
\label{fig:demospec}}
\end{figure*}

\begin{figure}
\epsscale{1.15}
\plottwo{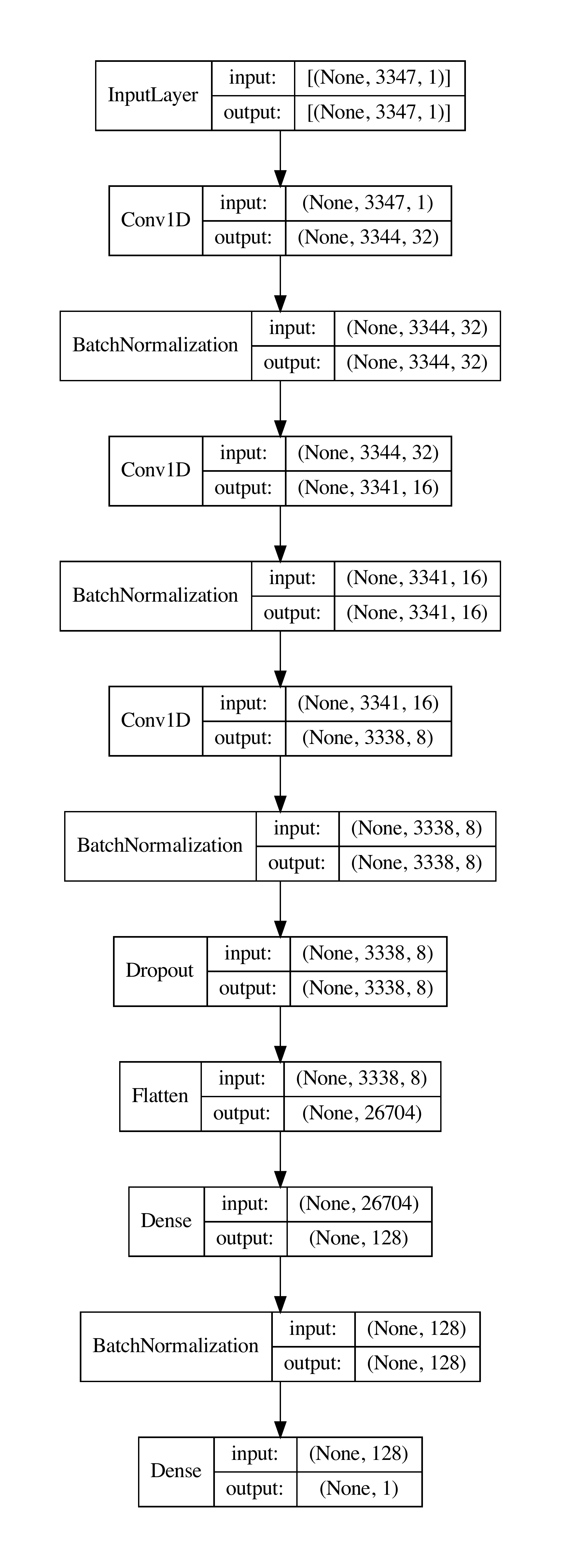}{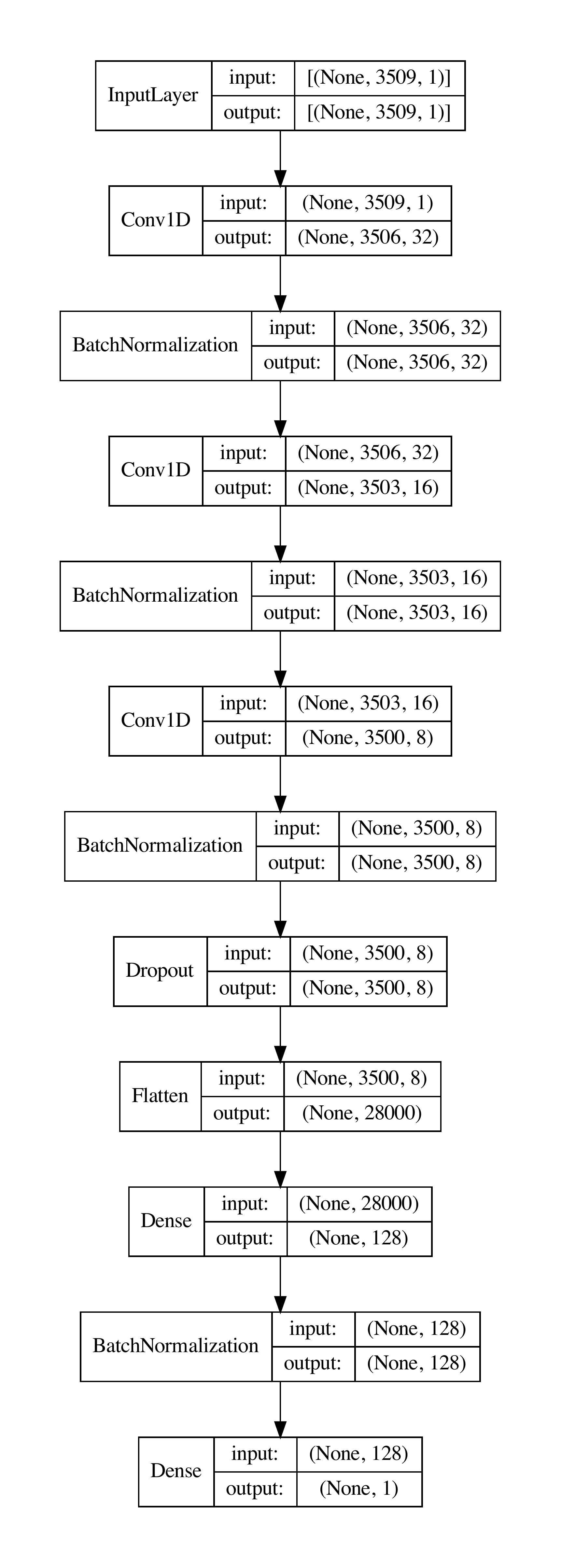}
\caption{The architectures of our CNN models for blue and red arms.
\label{fig:cnnstruct}}
\end{figure}

%\begin{figure}
%\epsscale{1.}
%\plotone{figv11/confusion_matrix.pdf}
%\caption{A schematic of the ideal confusion matrix.
%\label{fig:nnflow}}
%\end{figure}

\begin{figure}
\epsscale{1.2}
\plotone{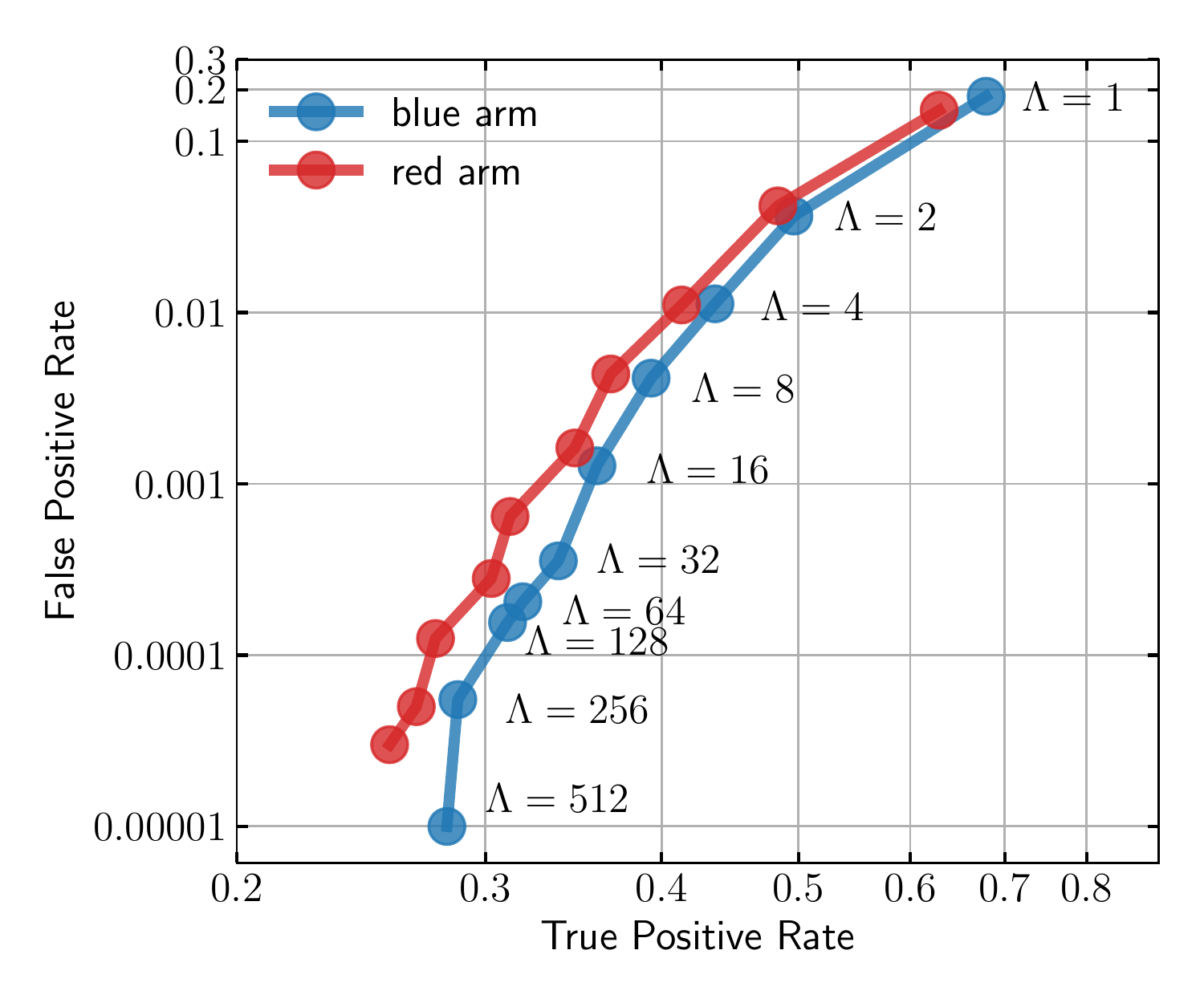}
\caption{The average False Positive Rates and True Positive Rates against $\Lambda$.
\label{fig:fpr_tpr}}
\end{figure}

\begin{figure*}
\epsscale{1.1}
%\plottwo{paperfigv16/Pb_Lambda_B.pdf}{paperfigv16/Pb_Lambda_R.pdf}
\plottwo{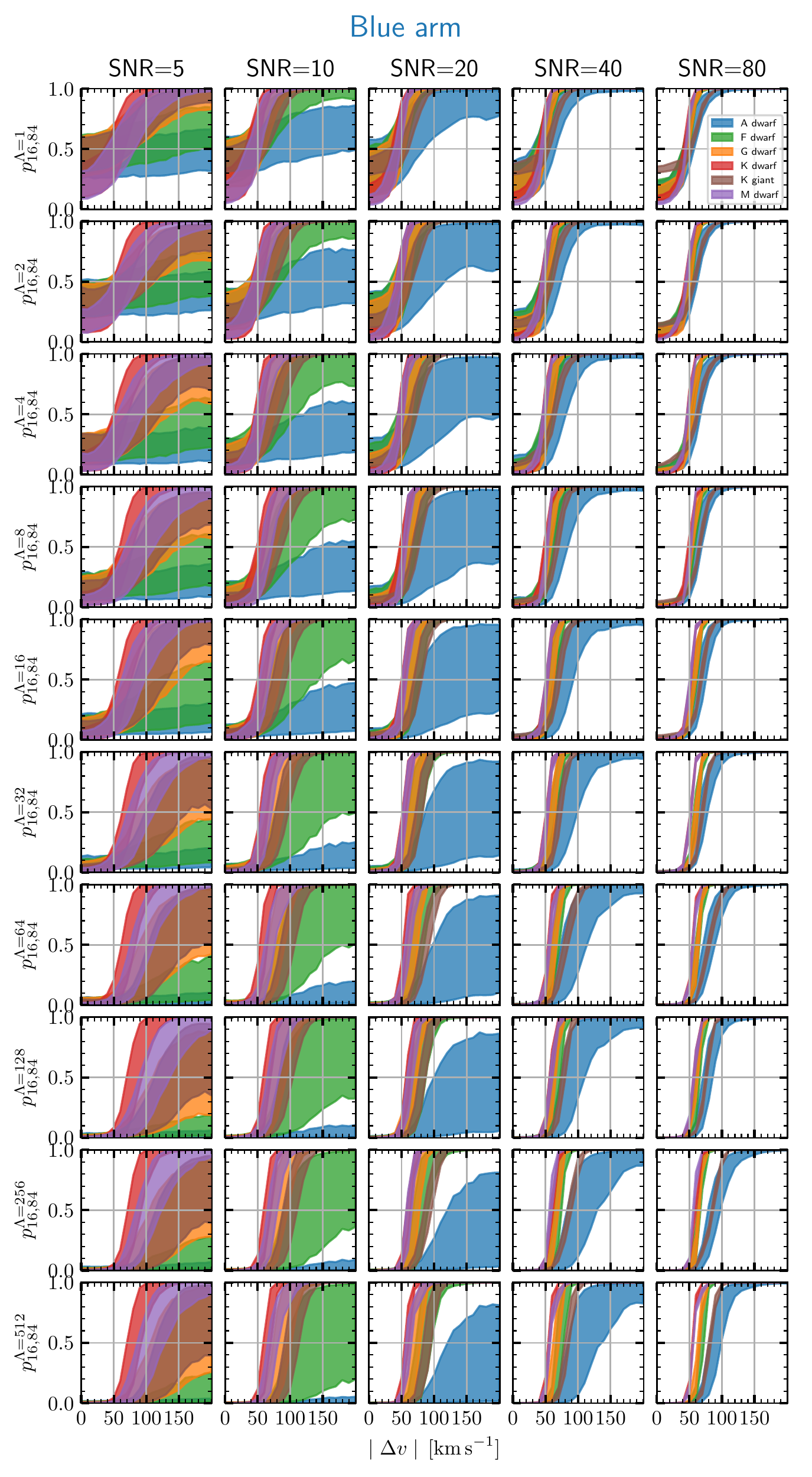}{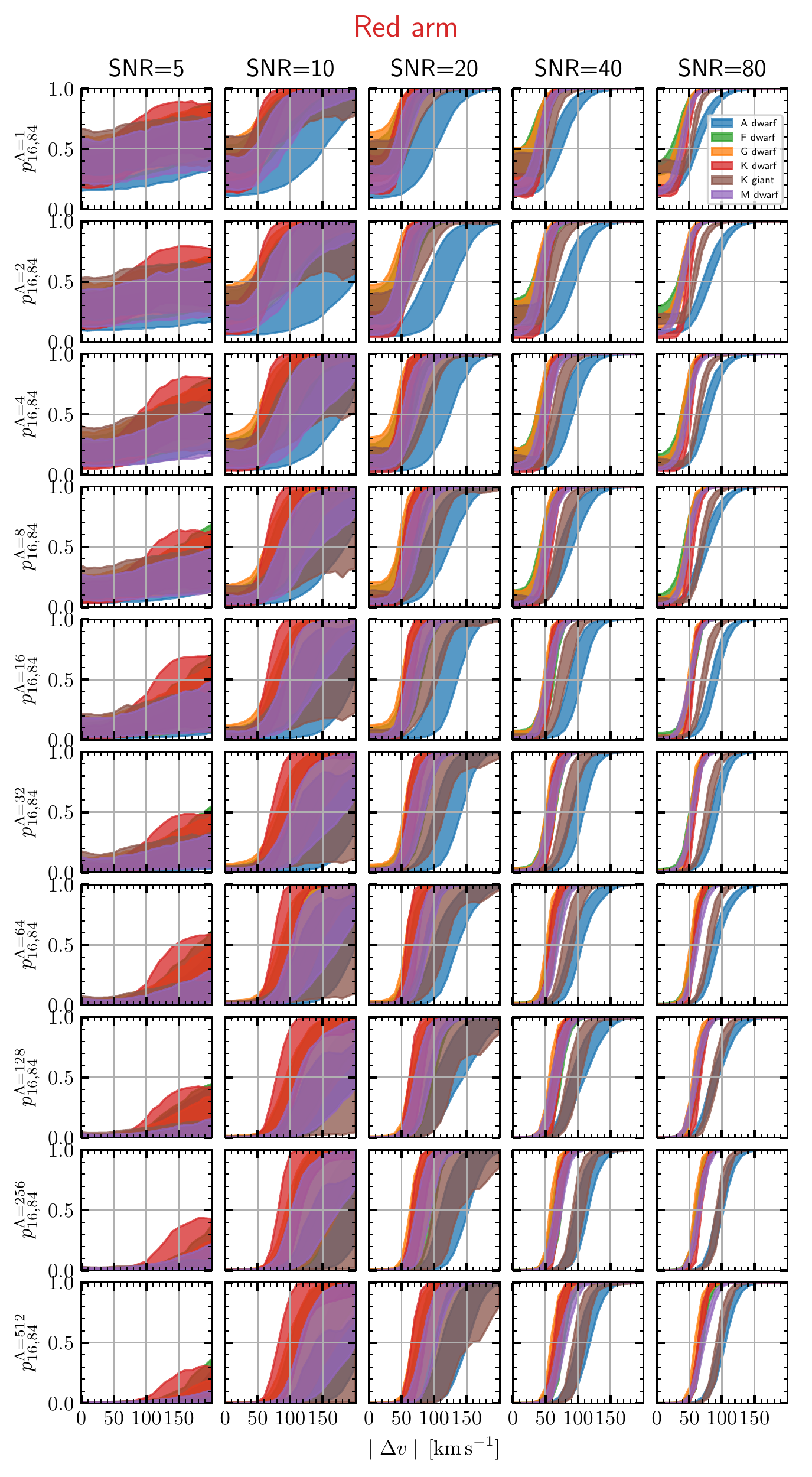}
\caption{Each panel in this figure shows the performance of our CNN models on mock main-sequence A-, F-, G-, K- and M-type twin binaries ($q=1$) at different S/N and $v_\mathrm{eq}\sin{i}$ values, assuming the two components in the binary system share the same parameters $T_\mathrm{eff}$, $\log{g}$, $\mathrm{[Fe/H]}$, $\mathrm{[\alpha/Fe]}$, and $v_\mathrm{eq}\sin{i}$.
\label{fig:Pb_snr_deltav}}
\end{figure*}

\begin{figure*}
\epsscale{1.1}
\plotone{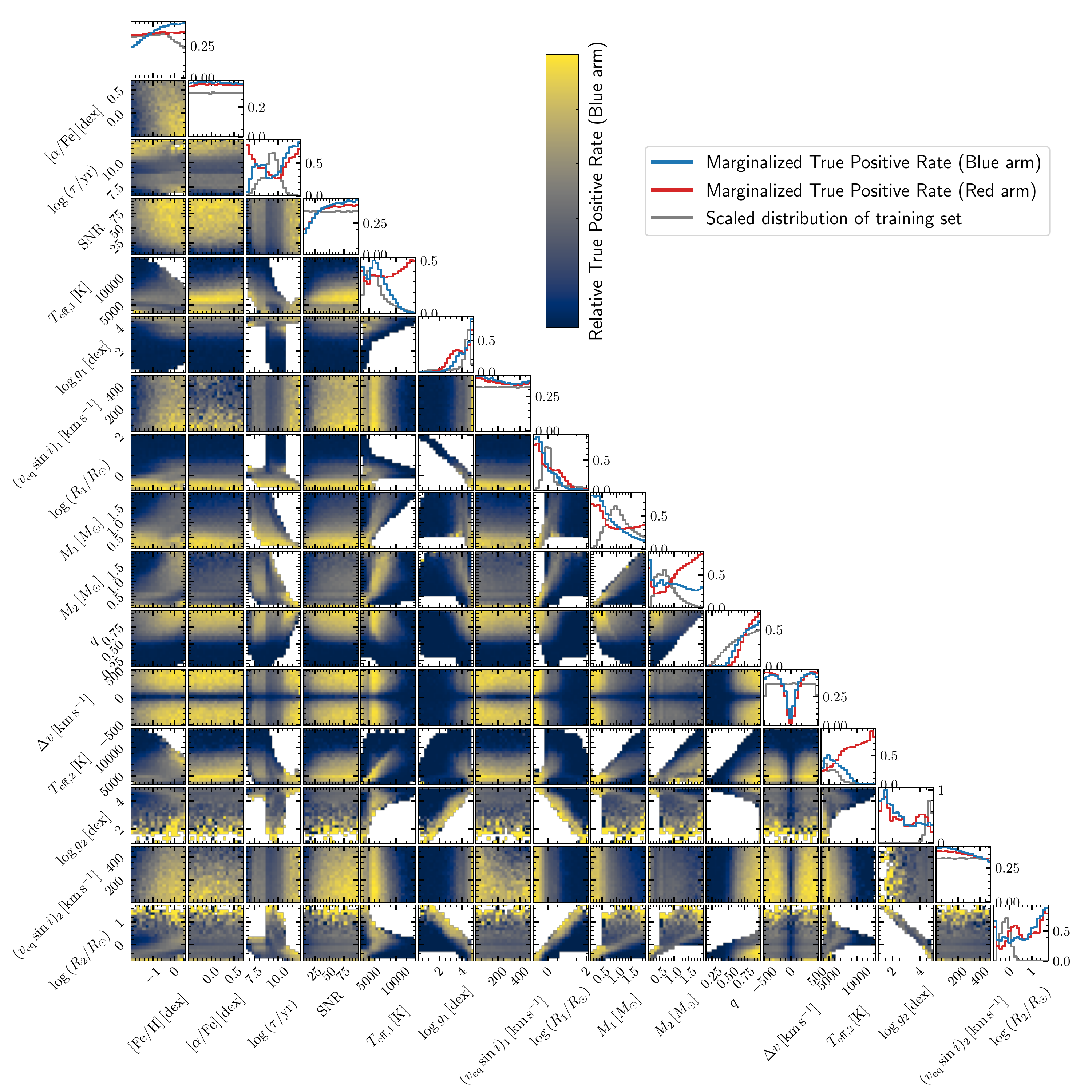}
\caption{The corner plot of average selection function of the models with $\Lambda=16$ for blue arm.
Those for red arm is similar so we do not show them in a separate figure. 
The diagonal panels show marginalized TPR (detectability) for blue/red arm in blue/red and the relative distribution of the training set in gray.
\label{fig:inference}}
\end{figure*}

\section{The discriminative model} \label{sec:model}

Neural networks entered stellar spectral classification and parameterization as an automated technique since several decades ago \citep{1994MNRAS.269...97V, 1997PASP..109..932B,1997MNRAS.292..157B}.
Recently, this technique and its variants, such as convolutional neural network (CNN), have become more and more popular in stellar spectroscopy for high computing performance.
They have been extensively used in both forward modeling \citep{2019ApJ...879...69T, 2019ApJS..245...34X}, backward modeling \citep{2019MNRAS.483.3255L, 2020A&A...644A.168G, 2020ApJ...891...23W}, and generative modeling \citep{2021FrASS...8...59Y, 2021ApJ...906..130O} of stellar spectra.
In this work, we use a CNN discriminative model to cope with the binary classification problem.

\subsection{The model architecture}
Our CNN model adopts the classic architecture of LeNet-5 \citep{1998PIEEELecun} but excludes the pooling layers to avoid losing spectral information.
The convolutional layers and fully connected layers use leaky ReLU (Rectified Linear Unit) activation function
\begin{equation}
\mathrm{ReLU}(x) = \begin{cases} x & x>=0  \\ 0.01 x & x<0 \end{cases}
\end{equation}
except the last layer uses a Sigmoid function
\begin{equation}
\mathrm{Sigmoid}(x) = \dfrac{1}{1+ e^{-x}}
\end{equation}
to restrict the output value in range [0, 1].
Therefore, the output could have a probabilistic interpretation.
BatchNormalization and Dropout layers are used as well to avoid gradient vanishing and overfitting problems.
The models are implemented using Keras\footnote{\url{https://keras.io/}}.
Figure \ref{fig:cnnstruct} shows our CNN architectures for the blue and red arms in left and right panels, respectively.

\subsection{The training strategy}
The training process minimizes binary cross-entropy (BCE) loss function
\begin{equation}
\begin{aligned}
	\ell_\mathrm{BCE}&(\hat{y}, y) = \\
	& - \frac{1}{N} \sum_{i=1}^n \left[ y_i\cdot \log{\hat{y}_i} + \Lambda \cdot (1-y_i) \cdot \log{(1-\hat{y}_i)} \right],
\end{aligned}
\end{equation}
where $\Lambda$ is the weight of negative samples over positive samples and can be interpreted as penalty.
Increasing $\Lambda$ suppresses false positives.
As a consequence, the probability of a spectrum being an SB2 depends on the training set and also the penalty parameter $\Lambda$.
Note that, our training sample is made up of pairs of positive and negative.
For example, the spectrum of a MS binary with a low mass ratio is very similar to a single star.
As this penalty $\Lambda$ goes high, the model is likely to classify the binary spectrum as negative.

We set a grid of $\Lambda$ values ([1, 2, 4, 8, 16, 32, 64, 128, 256, 512 ]), and for each value we train five models with different initialization.
Each model is optimized using the Adam optimizer with starting learning rate of 0.001.
The learning rate is divided by 5 once the loss on the validation set increases.
We use an early stopping strategy similar to \cite{2020A&A...644A.168G} to stop the training process automatically, i.e, training stops if the loss function increases for 10 consecutive epochs.
In each training process, the best model (with the lowest BCE) is saved using checkpoints.
Eventually, 50 models are trained and saved for blue and red arms, respectively.

\subsection{Estimator and uncertainty}
We denote the $i$th model with penalty $\Lambda$ by $\mathcal{M}^{\Lambda, i}$, where $i \in [1,2,3,4,5]$.
Theoretically, with the idea of Ensemble Learning (EL), averaging the results from five models with same $\Lambda$ gives a more accurate and robust estimation, i.e.,
\begin{equation} \label{eq:pavg}
	p^\Lambda = \frac{1}{5} \sum_{i=1}^5 \mathcal{M}^{\Lambda, i}(\boldsymbol{F}).
\end{equation}
In practice, we also take into account the flux error and use a Monte Carlo method to estimate $p^\Lambda$ and its error.
The $q$th percentile\footnote{Only in this case, $q$ denotes percentile rather than mass ratio.} of the Monte Carlo results can be evaluated via 
\begin{equation} \label{eq:pmc}
	p_q^\Lambda = \{\mathcal{M}^{\Lambda, i}(\boldsymbol{F}^j)\}_q,
\end{equation}
where $j$ is the serial number of Monte Carlo experiment, and the $\{\cdot\}_q$ is the $q$th percentile operator.

\subsection{False Positive Rate and True Positive Rate}
False Positive Rate (FPR) and True Positive Rate (TPR) can reflect contamination rate and detectability to some extent, respectively.
Their definitions are 
\begin{equation}
	FPR = \frac{N(\hat{y}>0.5 | y=0)}{N(y=0)},
\end{equation}
and
\begin{equation}
	TPR = \frac{N(\hat{y}>0.5 | y=1)}{N(y=1)},
\end{equation}
where $y$ is the truth and $\hat{y}$ is estimated binary probability.
Figure \ref{fig:fpr_tpr} shows the relation between mean FPR and mean TPR for specific $\Lambda$ evaluated on test set.
Generally, our models are more efficient in blue arm than red at a given $\Lambda$ since blue arm covers rich metal lines.
For both the blue and red arms, FPR and TPR decrease simultaneously with $\Lambda$.
This reflects that to avoid false positives (single stars classified to binaries), the models tend to drop more true binaries, particularly those whose secondary spectral component is not significant.
This is equivalent to adjusting threshold parameters in CCF method \citep[e.g.,][]{2017A&A...608A..95M}.
The choice of $\Lambda$ is a compromise between the FPR and TPR and is upon the specific scientific goal.
For the $\Lambda=16$ models the FPR is $\sim 0.001$ which is very low (40 false positives in the 40\,000 validation sample).
However, as the reviewer has kindly pointed out, the FPR is not the contamination rate of selected SB2 sample because the spectroscopic surveys are dominated by single stars and SB1s.
In this context, the FPR and TPR indicate the goodness of our models on our artificially constructed validation set.

\subsection{$\Lambda$ versus $\Delta v$} \label{sec:dv}
As an analogy to the CCF method, we also investigated the relationship between the predicted binary probability versus radial velocity separation ($\Delta v$) and S/N at different $\Lambda$.
Given S/N and $\Delta v$, we synthesize 2000 twin binary spectra ($\boldsymbol{\theta}_1=\boldsymbol{\theta}_2$, hence $q=1$) and add Gaussian noise following the way we synthesize our training set.
With our trained models $\mathcal{M}^{\Lambda,i}$, we summarize the predicted values by 16th and 84th percentiles of the Monte Carlo experiment.
The test twin binaries include
\begin{enumerate}
	\item A dwarf binary: $T_\mathrm{eff}=9000$ K, $\log{g}=4.5$ dex,
	\item F dwarf binary: $T_\mathrm{eff}=7000$ K, $\log{g}=4.5$ dex,
	\item G dwarf binary: $T_\mathrm{eff}=6000$ K, $\log{g}=4.5$ dex,
	\item K dwarf binary: $T_\mathrm{eff}=4500$ K, $\log{g}=4.8$ dex,
	\item K giant binary: $T_\mathrm{eff}=4500$ K, $\log{g}=1.5$ dex,
	\item M dwarf binary: $T_\mathrm{eff}=3500$ K, $\log{g}=4.8$ dex.
\end{enumerate}
The elemental abundances of the test twin binaries are assumed solar ($\mathrm{[Fe/H]}=0$ dex, $\mathrm{[\alpha/Fe]}=0$ dex) and the $v_\mathrm{eq}\sin{i}$ is fixed to $5\,\kms$.
In this test, the grid of S/N is (5, 10, 20, 40, 80), and the grid of $\Delta v$ is from 0  to 200 with a step of 10 $\kms$)

Figure \ref{fig:Pb_snr_deltav} shows the results of the test.
It indicates that our model is more efficient in recognizing F-, G-, K- and M-type dwarf binaries than A-type dwarf and K-type giant binaries which is due to the imbalance of the training set.
At the high S/N side, both blue and red arms show a jump at around $\Delta v \sim 50\,\kms$ because of the spectral resolution.
As $\Lambda$ increases, the jump moves toward larger $\Delta v$, meaning a large $\Lambda$ suppresses the identification of true binaries.

At modest S/N (20 to 40) which are typical for MRS spectra, we can detect most F-, G-, K- and M-type high mass ratio MS binaries as long as the $v_\mathrm{eq}\sin{i}$ is not too large and the observed $\Delta v$ (for K-type this threshold is lower).
The confusion region where the binary probability increases rapidly with increasing $\Delta v$ is from 30 to 100 km s$^{-1}$.
It is reasonable given that the MRS spectral resolution is $R\sim 7500$ and the resolution unit is $\dfrac{c}{R}\approx 40 \,\kms$.
For A- and perhaps earlier type stars, the detection of binaries is easier when using the red arm than the blue arm in low $v\sin{i}$ case.

\subsection{True Positive Rate as detectability}
It is interesting to have a glance at the behavior of the TRP against other parameters.
With the 40\,000 pairs in the test sample, we plot the marginalized TPR of our model ($\Lambda=16$, chosen based on Figure \ref{fig:fpr_tpr}) in blue arm in Figure \ref{fig:inference}.
In the diagonal panels, we also show the TPR at the red arm and the scaled distribution of our training set.
Several parameters are correlated strongly to the TPR of our model, for example, metallicity, the effective temperature of the primary star, and mass ratio.
Our model is efficient in detecting binaries whose primary star has a low effective temperature, a high metallicity, a high mass ratio ($q>0.7$), a large radial velocity separation ($\Delta v >50~\kms$, hence short period), and observed with high S/N, as expected.
The performance of the model at the red arm is similar to that in the blue arm but is more efficient in earlier type binaries. 
The imbalance of the training set also affects the performance.
For example, the inefficiency for red giant binaries is due to the lack of samples in the training set.
This figure is similar to Figure 5 in \cite{2010AJ....140..184M} but extended to many different dimensions, which helps to interpret the detectability of our model.

\begin{figure*}
\epsscale{1.1}
\plotone{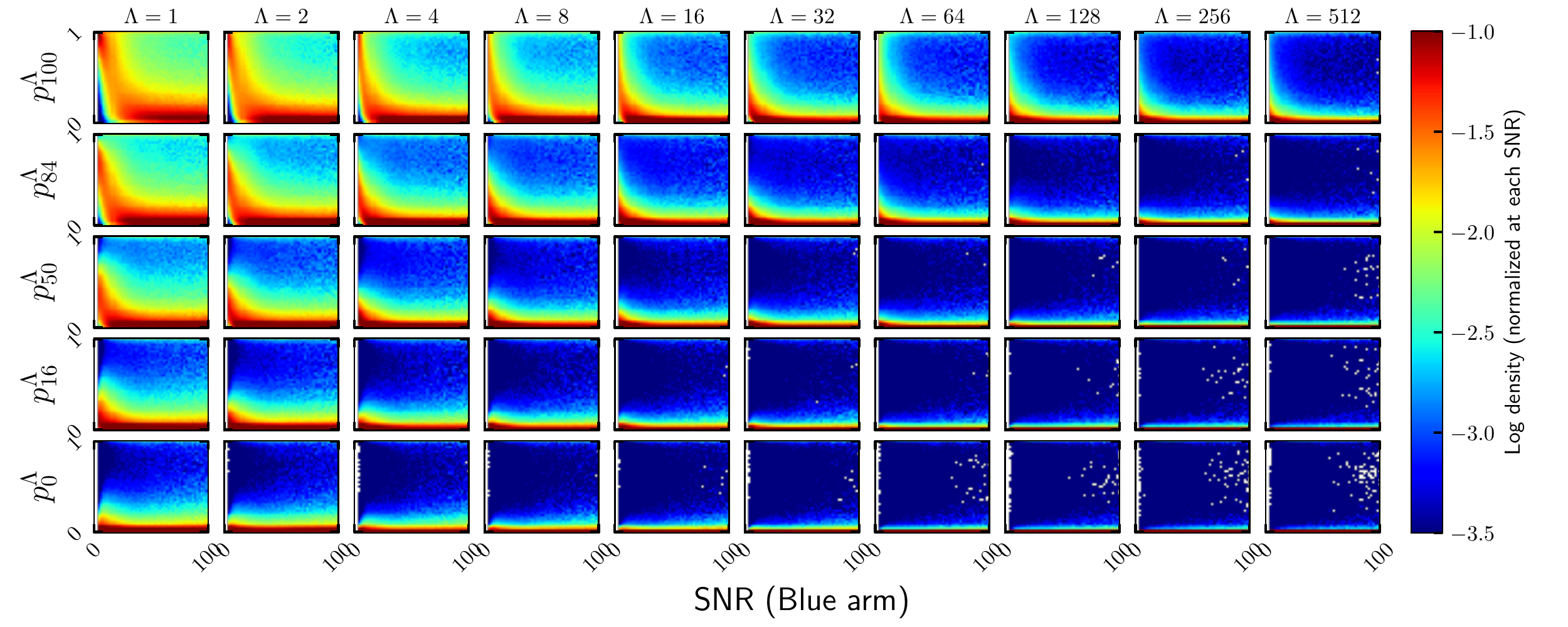}
\plotone{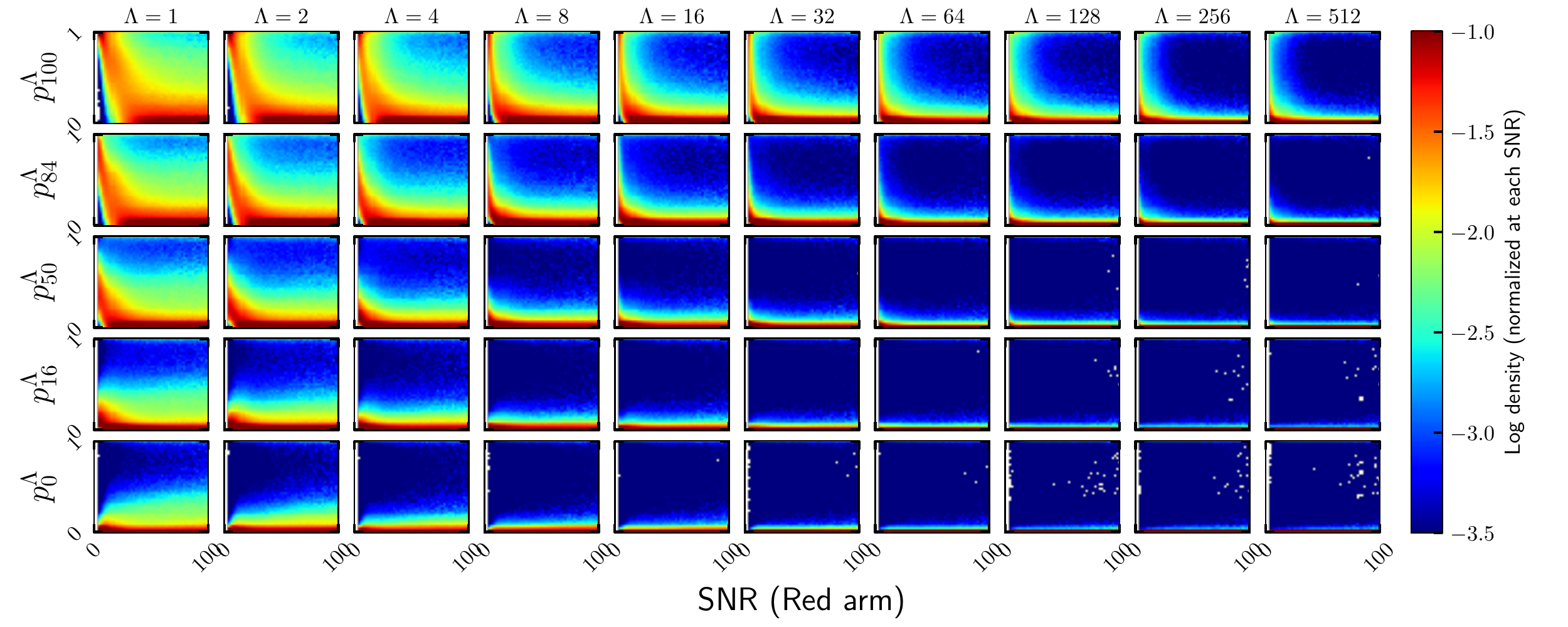}
\caption{The distribution of $p_q^{\Lambda}$ for the 5 million LAMOST MRS DR8 spectra with S/N$>5$, normalized for each S/N bin.
Both high $\Lambda$ and high $q$ values suppress false positives.
\label{fig:snr_pb}}
\end{figure*}

\begin{figure*}
\epsscale{1.1}
\plotone{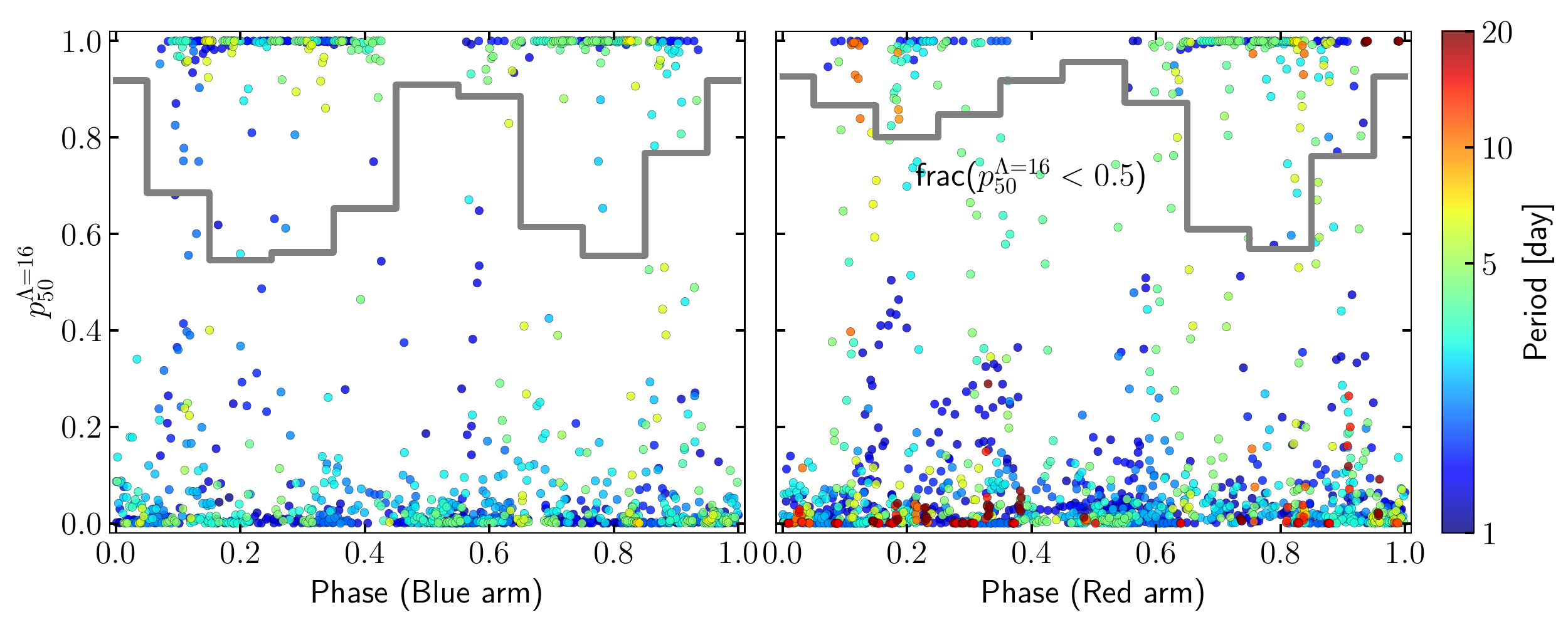}
\caption{Selected spectra of KEBC objects, including 1392 blue arm and 1492 red arm spectra.
Color codes the periods.
The thick gray lines represent the fraction of spectra with $p^{\Lambda=16}_{50}<0.5$.
0 and 1 are the phase corresponding to major eclipses.
\label{fig:phase_detectability}}
\end{figure*}

\begin{figure*}
\epsscale{1.2}
\plotone{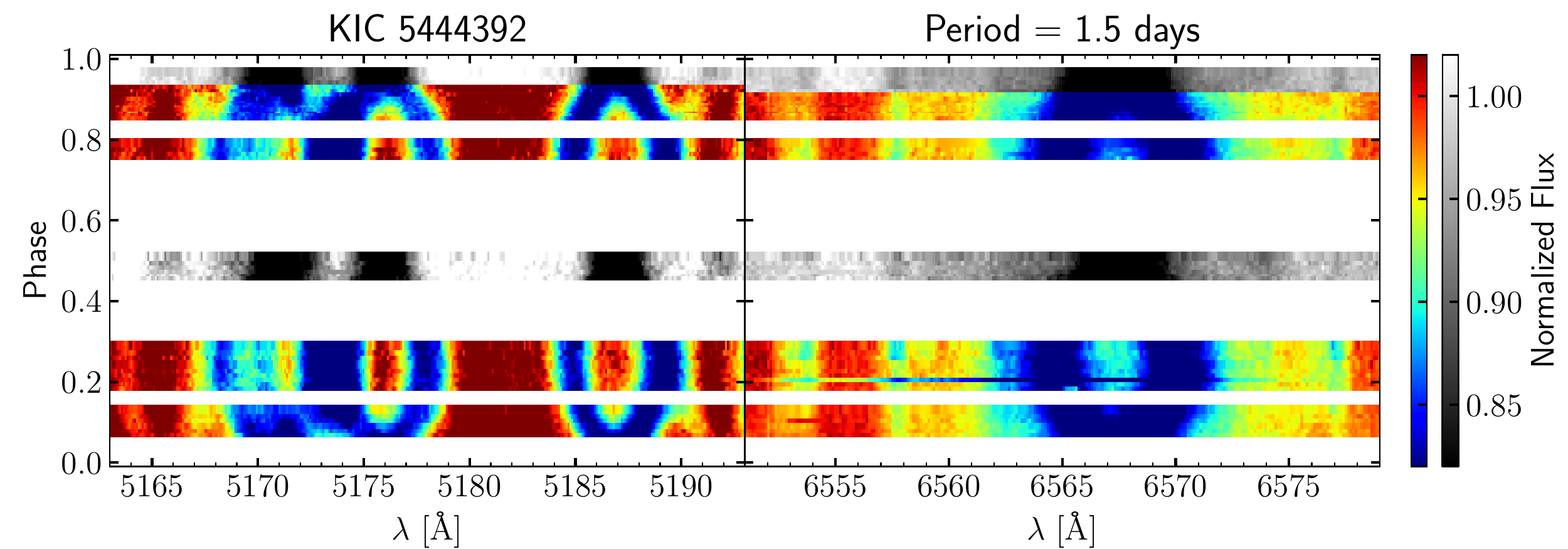}
\plotone{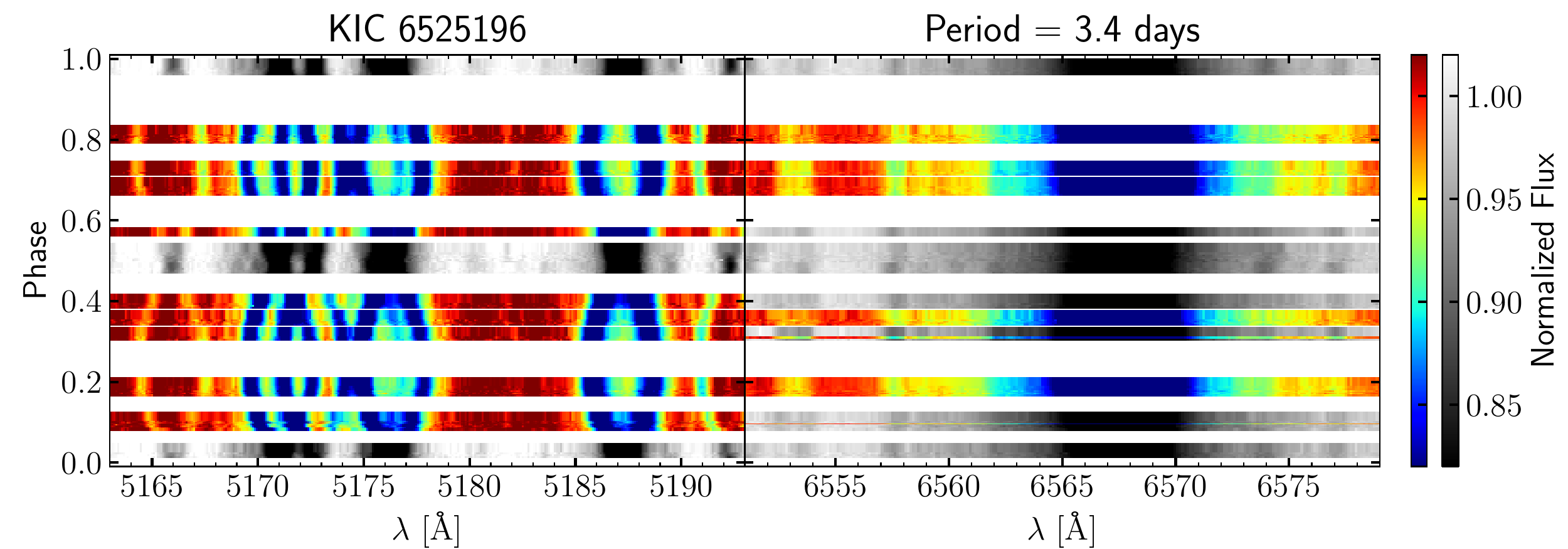}
\caption{Color coded spectra against folded phase derived from Kepler light curves for KIC5444392 (upper panel) and KIC6526196 (lower panel), respectively. The spectra with $p_{50}^{\Lambda=16}>0.5$ are shown using the "jet" color map while others are shown using the "gray" color map.
For briefness, only Mg $b$ and H$\alpha$ regions are shown. 
\label{fig:phase_spectra}}
\end{figure*}

\begin{figure*}
\epsscale{1.1}
\plotone{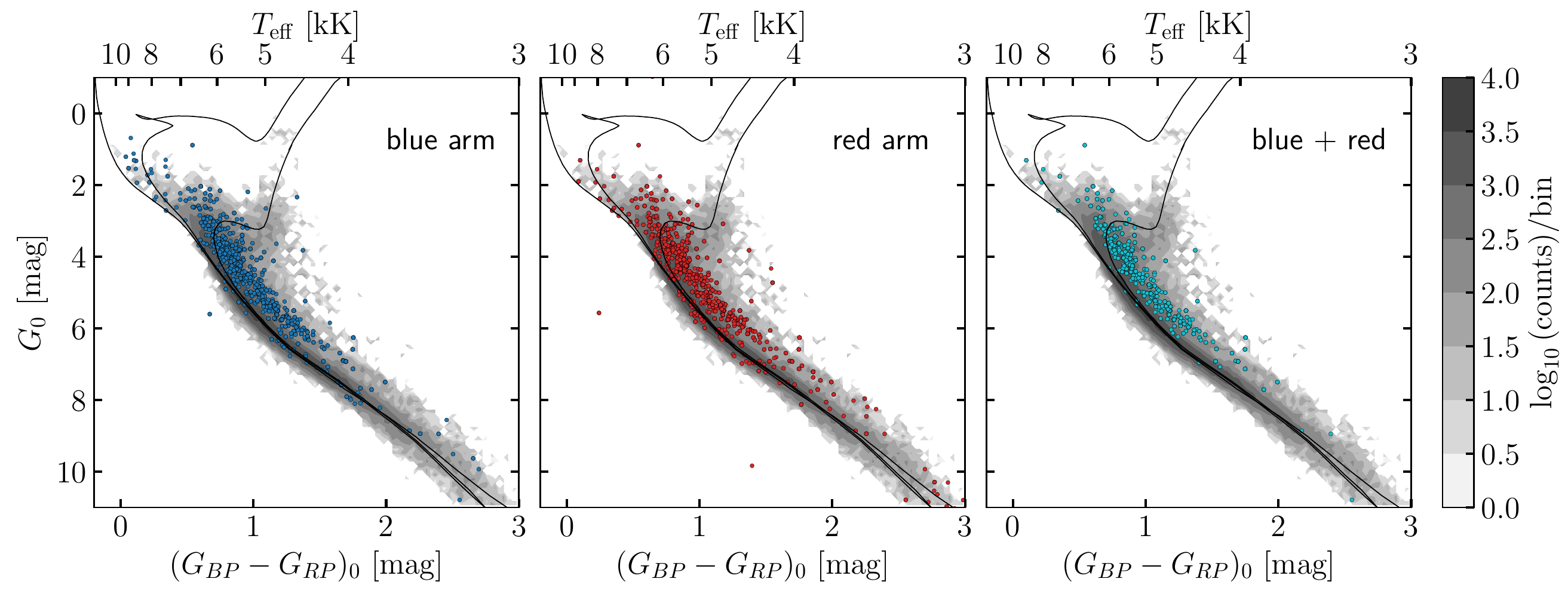}
\caption{The color-magnitude diagram of the LAMOST DR8 observed sources in solar neighborhood ($\varpi>3$).
$G_0$ is the absolute Gaia $G$ magnitude and $(G_{BP}-G_{RP})_0$ is the intrinsic color.
Gray contours show the distribution of all observations in LAMOST DR8 while colored dots have high binary probabilities. 
The left/middle/right panel shows the 2246/1876/1069 spectra with $p_0^{\Lambda=16}>0.5$ at blue/red/both arms.
The superposed isochrones have ages $\log{\tau}=$8.1, 8.9 and 9.7.
Approximate effective temperature is determined by interpolating the $\log\tau=8.1$ isochrone.
\label{fig:gaia_cmd}}
\end{figure*}

\begin{deluxetable*}{rlccl}
\tablewidth{0pt}
\tablenum{4}
\tablecaption{The 5 million binary probabilities and RVs obtained from the LAMOST MRS DR8 (v1.0). \label{tab:binarity}}
\tablehead{
\colhead{Index} &
\colhead{Label (FITS)} &
\colhead{Format} &
\colhead{Units} &
\colhead{Description}
}
\startdata
1 & obsid & Integer & --- & LAMOST observational ID (unique for each .fits file) \\
2 & lmjm & Integer & min & Local Modified Julian Minute \\
3 & bjdmid & Double & --- & Barycentric Julian Date of the middle of exposure \\
4 & lmjd & Integer & day & Local Modified Julian Day (only used for spectral naming) \\
5 & planid & String & --- & Plan ID \\
6 & spid & Short & --- & Spectrograph ID \\
7 & fiberid & Short & --- & Fiber ID \\
8 & ra & Double & deg & Right Ascension (J2000) \\
9 & dec & Double & deg & Declination (J2000) \\
10 & snr\_B & Float & --- & S/N ratio of blue arm \\
11 & snr\_R & Float & --- & S/N ratio of red arm \\
\hline
12 & rv\_obs\_B & Float & $\kms$ & RV measured from blue arm spectra \\
13 & rv\_obs\_err\_B & Float & $\kms$ & RV measurement error \\
14 & rv\_abs\_B & Float & $\kms$ & RV calibrated to Gaia eDR3 \\
15 & rv\_abs\_err\_B & Float & $\kms$ & total error of absolute RV \\
16 & rv\_teff\_B & Float & K & $T_\mathrm{eff}$ of the best template \\
17 & ccfmax\_B & Float & --- & CCF max value \\
18 & rv\_obs\_R & Float & $\kms$ & RV measured from red arm spectra \\
19 & rv\_obs\_err\_R & Float & $\kms$ & RV measurement error \\
20 & rv\_abs\_R & Float & $\kms$ & RV calibrated to Gaia eDR3 \\
21 & rv\_abs\_err\_R & Float & $\kms$ & total error of absolute RV \\
22 & rv\_teff\_R & Float & K & $T_\mathrm{eff}$ of the best template  \\
23 & ccfmax\_R & Float & --- & CCF max value  \\
\hline
24 & pb\_pct\_B\_8 & Float (5,) & --- & $p_q^\Lambda=8$ with $q=$0, 16, 50, 84 and 100 for blue arm \\
25 & pb\_pct\_B\_16 & Float (5,) & --- & $p_q^\Lambda=16$ with $q=$0, 16, 50, 84 and 100 for blue arm \\
26 & pb\_pct\_B\_32 & Float (5,) & --- & $p_q^\Lambda=32$ with $q=$0, 16, 50, 84 and 100 for blue arm \\
27 & specflag\_B & bool & --- & True for good spectra (S/N$>5$) \\
28 & npixbad\_B & Integer & --- & number of bad pixels \\
29 & pb\_pct\_R\_8 & Float (5,) & --- & $p_q^\Lambda=8$ with $q=$0, 16, 50, 84 and 100 for red arm \\
30 & pb\_pct\_R\_16 & Float (5,) & --- & $p_q^\Lambda=16$ with $q=$0, 16, 50, 84 and 100 for red arm \\
31 & pb\_pct\_R\_32 & Float (5,) & --- & $p_q^\Lambda=32$ with $q=$0, 16, 50, 84 and 100 for red arm \\
32 & specflag\_R & bool & --- & True for good spectra (S/N$>5$) \\
33 & npixbad\_R & Integer & --- & number of bad pixels \\
34 & bprp0 & Float & mag & Intrinsic color $(G_{BP}-G_{RP})_0$ \\
\enddata
\tablecomments{Table \ref{tab:binarity} is published in its entirety in the machine-readable (FITS) format. A portion is shown here for guidance regarding its form and content. File is available with the journal as well as at \url{https://github.com/hypergravity/paperdata}}
\end{deluxetable*}

\section{Results and validation} \label{sec:results}

\subsection{The spectroscopic binary probability catalog}

We measured the radial velocities for all the spectra ($S/N>5$) in LAMOST MRS DR8 v1.0\footnote{\url{http://www.lamost.org/dr8/}} following the method we proposed in \cite{2021ApJS..256...14Z}.
Then we shifted the spectra to the rest frame, interpolated to our pre-defined wavelength grid, removed cosmic rays roughly, and normalize them to continuum.
Applicating our model to observed data is straightforward at this stage.
To take into account the flux error, each spectrum is duplicated 20 times and added Gaussian random noise via its flux error.
For each $\Lambda$, the five trained model $\mathcal{M}^{\Lambda, i}$ predicts a binary probability for each spectrum in this Monte Carlo experiment.
This results in 100 binary probability values for one observed spectrum.
The 0th, 16th, 50th, 84th, and 100th percentiles are calculated to summarize the results.
To reduce the size of catalog, we only release the binary probabilities with $\Lambda=8$, $16$ and $32$ whose performance is reasonably good as we will see in the rest of this section.
The whole catalog consisting of RV measurements and binary probabilities for 5,893,382 spectra is described in Table \ref{tab:binarity} and is available at the journal as well as our Github website.
%\deleted{For the over 5 million spectra, the binary probability prediction takes 4 days on a Dell R740 workstation with two Intel Xeon Platinum 8270 CPUs @2.70GHz (26 cores each).}

Figure \ref{fig:snr_pb} shows the distribution of LAMOST MRS DR8 binary probabilities against S/N of spectra.
For visualization, we normalize counts in each S/N bin to eliminate the S/N distribution.
The decision boundary is clear at high $\Lambda$ while blurry at low $\Lambda$.
Combined with the results shown in Figure \ref{fig:fpr_tpr} and \ref{fig:Pb_snr_deltav}, we use the $16$ to suppress false positives for reliable results.
The users can choose their own values depending on specific scientific goals.

The identification of spectroscopic binaries always has a selection function depending on the adopted method, the instrumental properties, etc.
The experiments shown in Section \ref{sec:model} helps to investigate the performance on synthetic data.
However, it is difficult to calculate the 'true accuracy' of our model since we do not know the ground truth of observed objects.
Crossmatching our predicted binary probability and any other catalogs of spectroscopic binaries can only yield a qualitative comparison.
In the rest of this section, we show some qualitative results using identified eclipsing binaries and color-magnitude diagram, and present a cross-validation with APOGEE SB2 catalog.

\subsection{The Kepler Eclipsing Binaries} \label{sec:kebc}

The \textit{Kepler/K2} fields are extensively observed in MRS survey \citep{2020RAA....20..167F, 2020ApJS..251...15Z}.
Its targeting has taken into account the known eclipsing binaries identified from \textit{Kepler} ultra-precise photometry, namely the Kepler Eclipsing Binary Catalog \citep[hereafter KEBC,][]{2016AJ....151...68K} which compiles 2,922 well established eclipsing binaries (EBs).
Cross-matching the KEBC with our LAMOST MRS DR8 binary probability catalog yields 165 EBs with multiple observations (2813 spectra of whom at least one arm has $S/N>5$), among which the most intensively observed EB has 78 spectra.
%, among which 132 EBs have at least one spectra with usable quality $\mathrm{S/N_B}>5$ (1,544 spectra in total).%ootnote{Due to the instrumental efficiency, the signal-to-noise ratio of spectra is generally higher in the R band by $\sim 35\%$.}.
Folded phases are calculated from the BJD of the observation and Period and BJD$_0$ from the KEBC catalog.
Cutting $S/N>10$ and $1<P/ \mathrm{day}<10$, 1392/1492 spectra at blue/red arm are left.

Figure \ref{fig:phase_detectability} shows their relation between predicted binary probability ($p_{50}^{\Lambda=16}$) and the S/N of spectra at blue arm and red arm, respectively.
The sample shows clumps at low and high binary probability (0 and 1), and most high binary probability points occur in the non-eclipsing phase (0.1 to 0.4 and 0.6 to 0.9).
This is also reflected in the "W" shape fraction of objects with $p_{50}^{\Lambda=16}<0.5$ (gray thick line).
The binary probability is statistically highly correlated with phase, reflecting that our model is sensitive to the double-line features which are consistent with the results in Section \ref{sec:dv}.
%\deleted{We have checked it is not significantly correlated with period, because we eliminated long period EBs.}

As a demo, we present the spectra of two EBs (KIC 5444392 and KIC 6525196) ordered in folded phase and color-coded with normalized flux in Figure \ref{fig:phase_spectra}.
The spectra are divided into two groups.
Those with $p^{\Lambda=16}_{50}>0.5$ are shown with jet colormap and others with gray colormap.
The symmetric "8" shape Mg $b$ triplet and H$\alpha$ indicate their almost circular orbits.
The gray spectra, which are classified as 'single stars', consistently concentrate at eclipsing stages (phase $\sim 0, 0.5, 1.0$) despite only a few being misclassified.

\subsection{Color-magnitude diagram (CMD)}
The MS binary sequence gave us another perspective to check our results.
With criterion on Gaia eDR3 parallax \citep{2016A&A...595A...1G, 2021A&A...649A...1G} similar to \citep{2019MNRAS.490..550L}, i.e.,
\begin{enumerate}
	\item $\varpi>3$
	\item $\varpi / \sigma(\varpi)>5$
	\item ${\tt RUWE}<1.4$
	\item S/N$_B>10$ or S/N$_R>10$ 
\end{enumerate}
we arrive at 326,678/517,745 blue/red arm spectra, of which 2,246/1,876 are classified as binaries ($p_{0}^{\Lambda=16}>0.5$).
That our model detects more binaries in the blue arm is possibly due to more metal lines in the wavelength coverage \citep[see][]{2020ApJS..246....9Z,2020RAA....20...51Z}.
To calculate absolute magnitude, distance modulus is determined via
\begin{equation}
\mu = 5 \log_{10}{\frac{1000}{\varpi / \mathrm{arcsec}}} -5,
\end{equation}
and extinction is corrected using 3D dust map based on Gaia, Pan-STARRS1 and 2MASS \citep{ 2019ApJ...887...93G}.
The choice of minimum in Monte carlo results is to take into account the flux error.
In Figure \ref{fig:gaia_cmd}, we show the SB2 candidates selected from blue/red arm in the left/middle panel and their common subset in the right panel. 
The clear MS binary sequence in this CMD indicates that our model is sensitive to F-, G-, K-type MS binaries with high mass ratio.
This figure is very similar to those of spectroscopic binaries from GALAH \citep{2020A&A...638A.145T} and APOGEE \citep{2021AJ....162..184K} which are based on t-SNE and CCF.

\subsection{Cross-validation with APOGEE SB2s} \label{sec:cross_validation}
Cross-matching LAMOST MRS DR8 with APOGEE SB2s \citep{2021AJ....162..184K} results in 8152 spectra for 802 common SB2s, far more than with other SB2 catalogs \citep{2004A&A...424..727P, 2010AJ....140..184M, 2017A&A...608A..95M, 2018MNRAS.476..528E, 2020A&A...638A.145T},.
With the $p_{q=0}^{\Lambda=16}>0.5$ criteria, our model successfully identified 759 of 7023 blue arm spectra (10.8\%) and 501 of 8115 red arm spectra (6.2\%).
Reasons that hinder our model to discover the rest of SB2s are twofold.
From the observation side,
\begin{itemize}
	\item spectral resolution and S/N. The LAMOST MRS has $R\sim 7500$, lower than 22500 of APOGEE, thus detection of a binary requires a larger radial velocity separation (at least 50 $\kms$). As a consequence, the SB2s with period $>10$ days which are ubiquitous \citep{2010ApJS..190....1R} are mostly missing. And the typical S/N of LAMOST MRS spectra is 30, lower than the APOGEE survey, leading to a lower identification rate.
	\item wavelength. The detection limit of mass ratio $q$ for our model is around 0.7 which is higher than 0.4 reported by \citep{2018MNRAS.476..528E} for APOGEE. Typically, IR spectra have the advantage to discover an FGK-type MS SB2 since the luminosities of the two components are closer than in optical. Therefore, the IR SB2s are not necessarily SB2s in optical.
	\item observing phase. As shown in the KEBC experiment, an SB2, even with a high mass ratio, would not be identified if the spectra were obtained during or close to eclipses (see Section \ref{sec:kebc}). 
	\item spectral quality. The LAMOST pipeline has no routines on cosmic removal and quality check. Therefore, the released spectra often suffer from weird pixels. Although we have used a simple routine to remove the pixels with extreme values, we note that these pixels may affect our model prediction.
\end{itemize}
From the model side, the inefficiency in detecting some types of binaries, e.g., red giant binaries and early-type MS binaries, also leads to the low identification rate of APOGEE SB2s in our data to some extent.

\begin{figure*}
\epsscale{1.}
\plotone{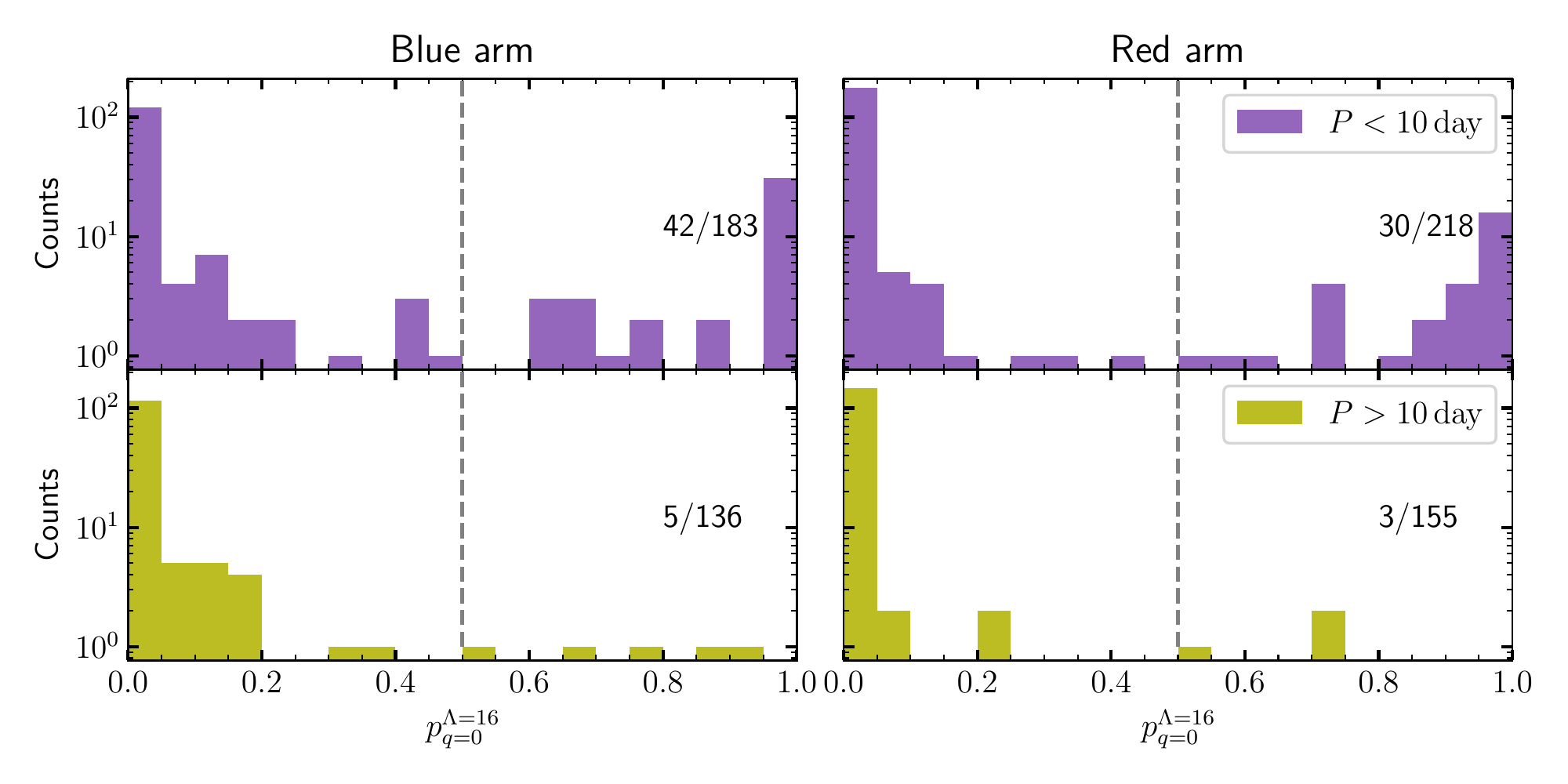}
\caption{The distribution of binary probabilities of the SB2s in APOGEE DR16\&17 \citep{2021AJ....162..184K} observed by LAMOST MRS.
The left/right panels show blue/red arms. Upper panels are for the very short period SB2s ($P<10$ days) while lower panels for SB2s with $P>10$ days.
\label{fig:recovery}}
\end{figure*}

In Figure \ref{fig:recovery}, we show the binary probability distribution of the subset of APOGEE SB2s with periods determined with the Joker \citep{2017ApJ...837...20P}.
The sample is split via their period $>10$ days or not.
The short period sample with $P>10$ days shows a 13-22\% identification rate, while the long period samples remain almost all unidentified in LAMOST MRS.
This indicates that the SB2s identified in the LAMOST MRS data are mostly with periods of $\lesssim 10$ days (with our model) and SB2s with periods $>10$ days are difficult to be identified.

\begin{figure*}
\epsscale{1.15}
\plotone{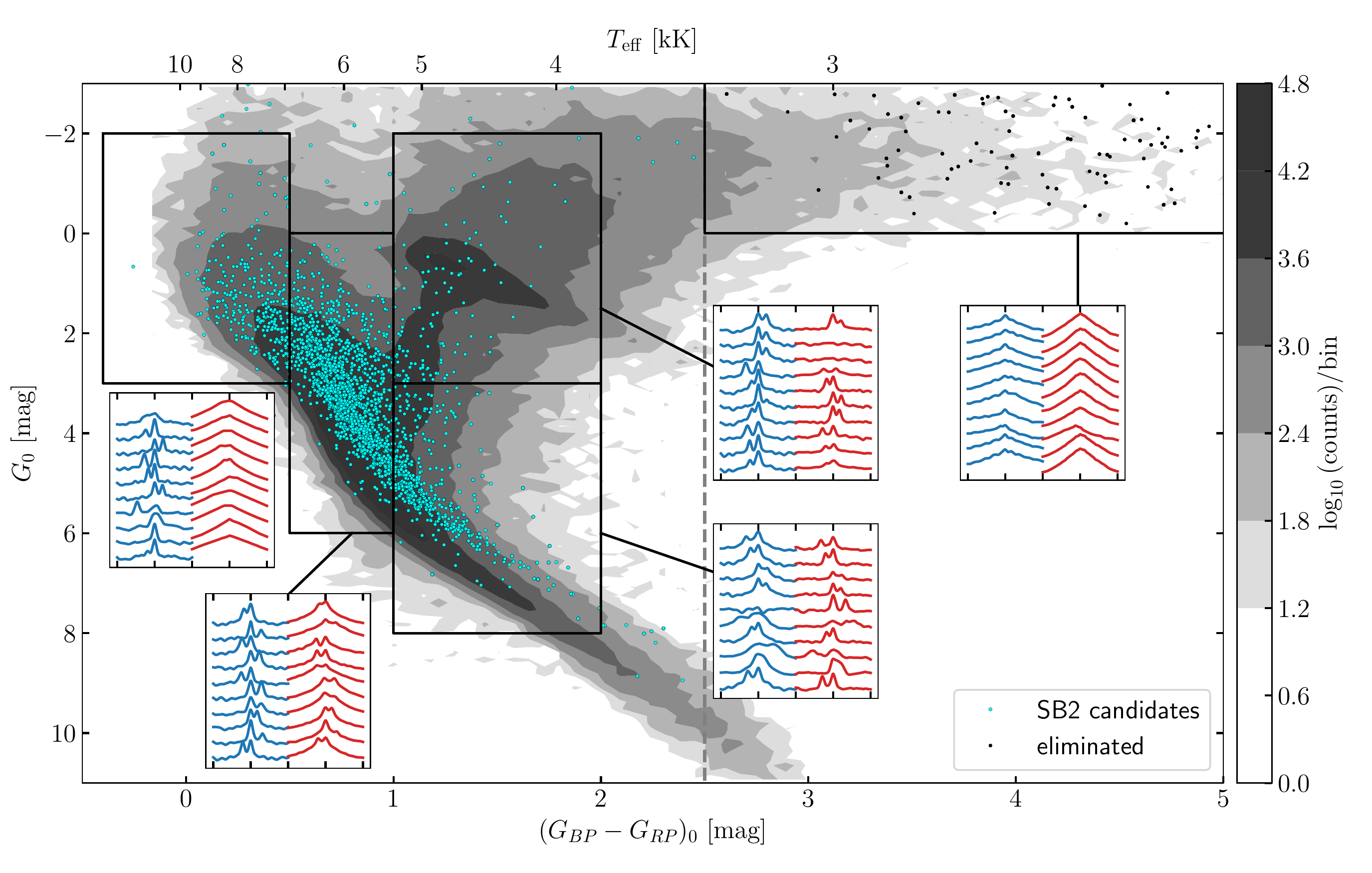}
\caption{The CMD of the SB2 candidates selected via high binary probability ($p^{\Lambda=16}_0>0.5$) in both arms.
The black dots represent the 116 eliminated objects and cyan dots are the final SB2 candidates.
Insets show the CCFs of randomly picked 10 stars from different regions in CMD.
Blue and red CCFs are for blue and red arms of the selected stars, respectively, in the $-600$ to $600\,\kms$ interval.
Note that the CCFs are corrected with their RVs (see Table \ref{tab:binarity}) to center the primary peaks.
The contour shows the distribution of all LAMOST MRS DR8 observations. 
\label{fig:sb2cmd}}
\end{figure*}

\begin{figure}
\epsscale{1.17}
\plotone{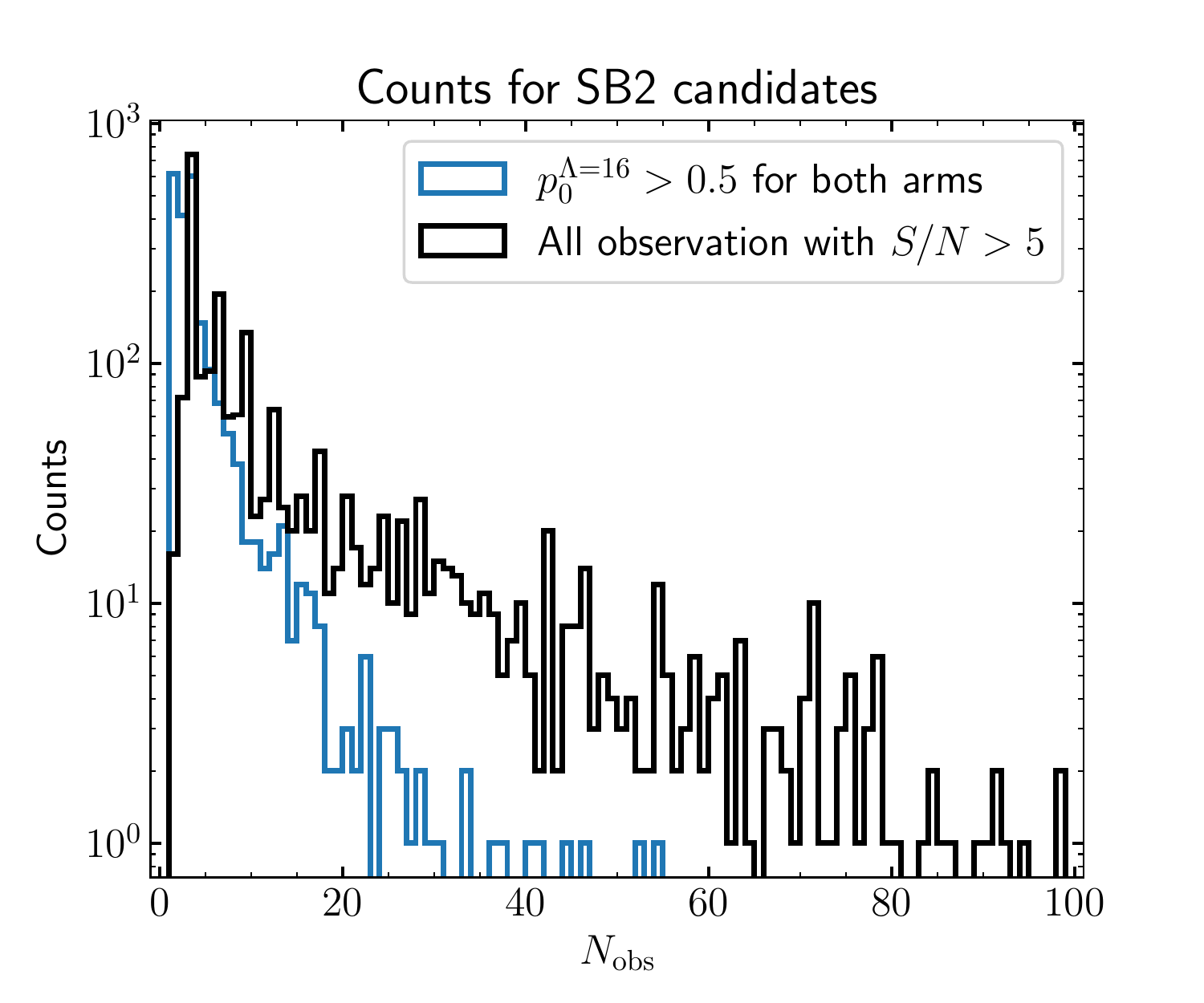}
\caption{
The selection
The counts of observations for the 2198 SB2 candidates with at least one epoch detected high binary probability ($p^{\Lambda=16}_0>0.5$) at both arms.
The blue shows the counts of observations with high binary probability, while the black shows the counts of all observations for these objects.
Bin width of $N_\mathrm{obs}$ is 1.
\label{fig:sb2counts}}
\end{figure}

\subsection{SB2 candidates}
At this stage, selecting a sample of spectroscopic binaries is straightforward.
Simply cutting $p^{\Lambda=16}_0>0.5$ and {\tt npixbad}$<100$ for both arms results in 8367 spectra for 2314 SB2 candidates.
However, we find that the identified SB2s are contaminated by AGB stars as we randomly pick out 10 stars from different regions in CMD and plot their CCFs in Figure \ref{fig:sb2cmd}.
This is reasonable if we recall that our training set has not included such cool giant stars.
At the high mass end, our sample is also possibly contaminated by fast rotators but not many.
For FGK MS and giant binaries, most CCFs show double peaks as expected.
Therefore, we additionally cut the $(G_{BP}-G_{RP})_0<2.5$ to eliminate those cool objects (364 spectra for 116 objects) which accounts for about 5\% of the whole sample.
The final sample consists of 8003 spectra for 2198 SB2 candidate.
The median of the numbers of observations is 6, while the median of the numbers of observations with high SB2 probabilities is 3.
1189 sources have at least 6 observations, which forms the basis of our following studies, for example, the determination of their atmospheric parameters and orbital parameters.
Figure \ref{fig:sb2counts} shows the numbers of observations for these sources.
The overall fraction of SB2 ($F_{SB2}$) for LAMOST MRS found in this work is less than 1\%, lower than other high-resolution surveys such as APOGEE \citep[3\% for dwarfs,][]{2021AJ....162..184K}.% for the reasons presented in Section \ref{sec:cross_validation}.

\section{Discussion} \label{sec:discussion}
The results of the above tests on synthetic data and observed data have proved the capability and feasibility of our model.
However, there are a few potential problems with it.
\begin{enumerate}
	\item The model is automatic and efficient in detecting the FGK-type MS binaries rather than others (e.g., red giant binaries as we see from Figure \ref{fig:Pb_snr_deltav}) due to the imbalance of our training set. Therefore, the objects with high predicted binary probability are biased in spectral types and etc. Other types of binaries, such as white dwarf--Main sequence binaries or white dwarf--white dwarf binaries, have not been considered yet in this work. 
	\item There could be some inconsistencies between the results from the blue arm and red arm (Figure \ref{fig:phase_spectra}). Combining two arms can improve the reliability of results (Figure \ref{fig:gaia_cmd}) but may result in the loss of a number of true SB2s. A direct way to improve the consistency is to train the models for the two arms simultaneously. 
	\item Like other classification methods based on machine learning \citep[e.g.,][]{2020A&A...638A.145T}, the true TPR can not be evaluated using any existing dataset. Currently, the overall $F_{SB2}$ for LAMOST MRS is low compared to other high-resolution spectroscopic surveys due to various reasons. The identified SB2s are mainly those with very short periods $P\lesssim10$ days.
	\item The results can be affected by other potential problems, e.g., bad pixels or emission lines in spectra. Models based on neural networks are sensitive to those spectral features which are absent in the training set. In this regard, the CCF method is more robust.
	\item Random line of sight alignments of unrelated stars \citep[see][]{2021AJ....162..184K} are not considered in this work but are potential considering the typical seeing of LAMOST \citep[$\sim 3$ arcsec,][]{2015PASP..127.1292Z}. 
\end{enumerate} 
%\subsection{Choosing a proper model}
We note that the choice of $\Lambda$ and $q$ determines the selection function of the model and results in different FPR and TPR.
The models with low $\Lambda$ values may report false positives but also find more true positives, whereas models with high $\Lambda$ deliver us a reliable SB2 sample with low TPR.
Therefore, it balances the tradeoff between TPR and FPR and relies on specific science goals.
For example, our low $\Lambda$ model has been used in the LAMOST-TD survey \citep{2021arXiv210903149W} to prompt possible SB2 candidates.
It serves as an aid parallel to manual inspection and works well.

Last but not least, the FPR and TPR are calculated based on our mock data set which in many dimensions has a distribution covering a very wide range while in some dimensions narrow.
For example, the upper bound of the $v \sin{i}$ is set to 500 $\kms$ which is rarely seen among FGK-type stars, while long-period binaries rarely occur in our sample due to the uniform distribution of $\Delta v$.
Therefore, our estimated FPR (e.g., 0.001 for $\Lambda=16$) and TPR (e.g., 30\% for $\Lambda=16$) should be referred to with caution when applying our model to real data whose distribution is different.
%On the other hand, the TPR=30\% is optimistic since our synthetic data set emphasizes on short period binaries.
%Long period SB2s are unlikely to be detected given our spectral resolution.

\section{Conclusions and prospects} \label{sec:conclusions}

We developed a model based on Convolutional neural network (CNN) to automatically distinguish double-lined spectroscopic binaries (SB2s) from single stars/single-lined spectroscopic binaries (SB1s) using single exposure spectra in a probabilistic approach.
This model is trained on a large set of synthetic spectra with $R=7\,500$ based on the ATLAS9 model and MIST stellar evolutionary model.
It reaches a reasonable high accuracy given a proper penalty on the negative sample.
It can efficiently detect F-, G-, K-type MS binaries from LAMOST MRS spectra with large radial velocity separation (hence short period, $\lesssim 10$ days) and high mass ratio ($q>0.7$).
The selection function is inherited from the imbalance of the training sample.
%Its performance is as good as the traditional CCF method, and the probabilistic essence fits in the hierarchical modeling of binaries.

This model has been tested on a synthetic dataset and shows novel performance with $\Lambda=16$.
The selection function is estimated as well.
We have validated the model using Kepler eclipsing binaries and it shows an interesting binary probability-phase relation.
The high binary probabilities consistently occur at non-eclipsing phases in our statistics.
%Two well-observed demo EBs also
We also validate the high binary probability objects with Gaia CMD of nearby stars and show a clear MS binary sequence, meaning that our model is efficient in detecting binaries with high mass ratio.
Cross-matching with APOGEE SB2s reveals the fact that for LAMOST MRS the majority of SB2s detected have a very short period ($\lesssim10$ days).
Although quantitative assessment of accuracy is not flexible like for other machine learning classification methods, the inspection on the CCFs of randomly picked SB2 candidates show that the major contamination ($\sim$5\%) comes from AGB stars whose features are very wide.
Since our training set does not include any such cool giants, this contamination is acceptable.
And we claim that our model is reasonably reliable.

This model leads to an immediate application on the LAMOST MRS survey, which contains over 5 million single exposure medium-resolution spectra with S/N$>5$.
We build a catalog of 2198 SB2 candidates which has at least one spectra with high binary probability in both blue and red arms.
With multi-epoch spectra from the LAMOST MRS survey, we will analyze their atmospheric parameters and orbital properties in our next paper.

\acknowledgments
BZ thanks Prof. Hans-Walter Rix for the generous support for the visit to MPIA.
The authors thank the reviewer for useful comments to the manuscript.
%BZ thanks Prof. Hans-Walter Rix because BZ's understandings of the stellar evolutionary tracks and much of the code in this work was established during the days when BZ was visiting MPIA.
%BZ thanks Zhao-Zhou Li for sharing several useful {\tt python} wheels which simplified this work.
%BZ also thanks his mother for preparing his favorite food everyday during the epidemic.
%JNF acknowledges the support from the National Natural Science Foundation of China (NSFC) through the grants 11833002 and 11673003.
JNF acknowledges the support from the National Natural Science Foundation of China (NSFC) through the grants 11833002, 12090040 and 12090042.
HT is supported by Beijing Natural Science Foundation with grant No. 1214028, and partly supported by National Natural Science Foundation of China with grant No.U2031143.
%YHH acknowledges the support by Key Research Program of Frontier Sciences, CAS, Grant NO. QYZDY-SSW-SLH007.
WZ is supported by the Fundamental Research Funds for the central universities and NSFC 11903005.

The LAMOST FELLOWSHIP is supported by Special Funding for Advanced Users, budgeted and administrated by Center for Astronomical Mega-Science, Chinese Academy of Sciences (CAMS). This work is supported by Cultivation Project for LAMOST Scientific Payoff and Research Achievement of CAMS-CAS.

Guoshoujing Telescope (the Large Sky Area Multi-Object Fiber Spectroscopic Telescope LAMOST) is a National Major Scientific Project built by the Chinese Academy of Sciences. Funding for the project has been provided by the National Development and Reform Commission. LAMOST is operated and managed by the National Astronomical Observatories, Chinese Academy of Sciences.


\vspace{5mm}
\facilities{LAMOST}


\software{%regli\citep{regli},
berliner\citep{2020zndo...4381163Z},
laspec\citep{2020zndo...4381155Z},
astropy \citep{2013A&A...558A..33A},
scikit-learn \citep{2012arXiv1201.0490P},
dustmaps \citep{2018JOSS....3..695M}}
% \software{astropy \citep{2013A&A...558A..33A},  
%           Cloudy \citep{2013RMxAA..49..137F}, 
%           SExtractor \citep{1996A&AS..117..393B}
%           }



\newpage

\bibliography{cnnsb.bib}

\begin{thebibliography}{}
\expandafter\ifx\csname natexlab\endcsname\relax\def\natexlab#1{#1}\fi
\providecommand{\url}[1]{\href{#1}{#1}}
\providecommand{\dodoi}[1]{doi:~\href{http://doi.org/#1}{\nolinkurl{#1}}}
\providecommand{\doeprint}[1]{\href{http://ascl.net/#1}{\nolinkurl{http://ascl.net/#1}}}
\providecommand{\doarXiv}[1]{\href{https://arxiv.org/abs/#1}{\nolinkurl{https://arxiv.org/abs/#1}}}

\bibitem[{{Abt} \& {Levy}(1976)}]{1976ApJS...30..273A}
{Abt}, H.~A., \& {Levy}, S.~G. 1976, \apjs, 30, 273, \dodoi{10.1086/190363}

\bibitem[{{Allende Prieto} {et~al.}(2018){Allende Prieto}, {Koesterke},
  {Hubeny}, {Bautista}, {Barklem}, \& {Nahar}}]{2018A&A...618A..25A}
{Allende Prieto}, C., {Koesterke}, L., {Hubeny}, I., {et~al.} 2018, \aap, 618,
  A25, \dodoi{10.1051/0004-6361/201732484}

\bibitem[{{Astropy Collaboration} {et~al.}(2013){Astropy Collaboration},
  {Robitaille}, {Tollerud}, {Greenfield}, {Droettboom}, {Bray}, {Aldcroft},
  {Davis}, {Ginsburg}, {Price-Whelan}, {Kerzendorf}, {Conley}, {Crighton},
  {Barbary}, {Muna}, {Ferguson}, {Grollier}, {Parikh}, {Nair}, {Unther},
  {Deil}, {Woillez}, {Conseil}, {Kramer}, {Turner}, {Singer}, {Fox}, {Weaver},
  {Zabalza}, {Edwards}, {Azalee Bostroem}, {Burke}, {Casey}, {Crawford},
  {Dencheva}, {Ely}, {Jenness}, {Labrie}, {Lim}, {Pierfederici}, {Pontzen},
  {Ptak}, {Refsdal}, {Servillat}, \& {Streicher}}]{2013A&A...558A..33A}
{Astropy Collaboration}, {Robitaille}, T.~P., {Tollerud}, E.~J., {et~al.} 2013,
  \aap, 558, A33, \dodoi{10.1051/0004-6361/201322068}

\bibitem[{{Bailer-Jones}(1997)}]{1997PASP..109..932B}
{Bailer-Jones}, C.~A.~L. 1997, \pasp, 109, 932, \dodoi{10.1086/133962}

\bibitem[{{Bailer-Jones} {et~al.}(1997){Bailer-Jones}, {Irwin}, {Gilmore}, \&
  {von Hippel}}]{1997MNRAS.292..157B}
{Bailer-Jones}, C. A.~L., {Irwin}, M., {Gilmore}, G., \& {von Hippel}, T. 1997,
  \mnras, 292, 157, \dodoi{10.1093/mnras/292.1.157}

\bibitem[{{Borucki} {et~al.}(2010){Borucki}, {Koch}, {Basri}, {Batalha},
  {Brown}, {Caldwell}, {Caldwell}, {Christensen-Dalsgaard}, {Cochran},
  {DeVore}, {Dunham}, {Dupree}, {Gautier}, {Geary}, {Gilliland}, {Gould},
  {Howell}, {Jenkins}, {Kondo}, {Latham}, {Marcy}, {Meibom}, {Kjeldsen},
  {Lissauer}, {Monet}, {Morrison}, {Sasselov}, {Tarter}, {Boss}, {Brownlee},
  {Owen}, {Buzasi}, {Charbonneau}, {Doyle}, {Fortney}, {Ford}, {Holman},
  {Seager}, {Steffen}, {Welsh}, {Rowe}, {Anderson}, {Buchhave}, {Ciardi},
  {Walkowicz}, {Sherry}, {Horch}, {Isaacson}, {Everett}, {Fischer}, {Torres},
  {Johnson}, {Endl}, {MacQueen}, {Bryson}, {Dotson}, {Haas}, {Kolodziejczak},
  {Van Cleve}, {Chandrasekaran}, {Twicken}, {Quintana}, {Clarke}, {Allen},
  {Li}, {Wu}, {Tenenbaum}, {Verner}, {Bruhweiler}, {Barnes}, \&
  {Prsa}}]{2010Sci...327..977B}
{Borucki}, W.~J., {Koch}, D., {Basri}, G., {et~al.} 2010, Science, 327, 977,
  \dodoi{10.1126/science.1185402}

\bibitem[{{Choi} {et~al.}(2016){Choi}, {Dotter}, {Conroy}, {Cantiello},
  {Paxton}, \& {Johnson}}]{2016ApJ...823..102C}
{Choi}, J., {Dotter}, A., {Conroy}, C., {et~al.} 2016, \apj, 823, 102,
  \dodoi{10.3847/0004-637X/823/2/102}

\bibitem[{{Cui} {et~al.}(2012){Cui}, {Zhao}, {Chu}, {Li}, {Li}, {Zhang}, {Su},
  {Yao}, {Wang}, {Xing}, {Li}, {Zhu}, {Wang}, {Gu}, {Luo}, {Xu}, {Zhang},
  {Liu}, {Zhang}, {Yang}, {Cao}, {Chen}, {Chen}, {Chen}, {Chen}, {Chu}, {Feng},
  {Gong}, {Hou}, {Hu}, {Hu}, {Hu}, {Jia}, {Jiang}, {Jiang}, {Jiang}, {Jin},
  {Li}, {Li}, {Li}, {Liu}, {Liu}, {Lu}, {Mao}, {Men}, {Qi}, {Qi}, {Shi},
  {Tang}, {Tao}, {Wang}, {Wang}, {Wang}, {Wang}, {Wang}, {Wang}, {Wang},
  {Wang}, {Wang}, {Wang}, {Wang}, {Wang}, {Xu}, {Xu}, {Yang}, {Yu}, {Yuan},
  {Yuan}, {Zhai}, {Zhang}, {Zhang}, {Zhang}, {Zhao}, {Zhou}, {Zhou}, {Zhu}, \&
  {Zou}}]{2012RAA....12.1197C}
{Cui}, X.-Q., {Zhao}, Y.-H., {Chu}, Y.-Q., {et~al.} 2012, Research in Astronomy
  and Astrophysics, 12, 1197, \dodoi{10.1088/1674-4527/12/9/003}

\bibitem[{{De Silva} {et~al.}(2015){De Silva}, {Freeman}, {Bland-Hawthorn},
  {Martell}, {de Boer}, {Asplund}, {Keller}, {Sharma}, {Zucker}, {Zwitter},
  {Anguiano}, {Bacigalupo}, {Bayliss}, {Beavis}, {Bergemann}, {Campbell},
  {Cannon}, {Carollo}, {Casagrande}, {Casey}, {Da Costa}, {D'Orazi}, {Dotter},
  {Duong}, {Heger}, {Ireland}, {Kafle}, {Kos}, {Lattanzio}, {Lewis}, {Lin},
  {Lind}, {Munari}, {Nataf}, {O'Toole}, {Parker}, {Reid}, {Schlesinger},
  {Sheinis}, {Simpson}, {Stello}, {Ting}, {Traven}, {Watson}, {Wittenmyer},
  {Yong}, \& {{\v{Z}}erjal}}]{2015MNRAS.449.2604D}
{De Silva}, G.~M., {Freeman}, K.~C., {Bland-Hawthorn}, J., {et~al.} 2015,
  \mnras, 449, 2604, \dodoi{10.1093/mnras/stv327}

\bibitem[{{Deng} {et~al.}(2012){Deng}, {Newberg}, {Liu}, {Carlin}, {Beers},
  {Chen}, {Chen}, {Christlieb}, {Grillmair}, {Guhathakurta}, {Han}, {Hou},
  {Lee}, {L{\'e}pine}, {Li}, {Liu}, {Pan}, {Sellwood}, {Wang}, {Wang}, {Yang},
  {Yanny}, {Zhang}, {Zhang}, {Zheng}, \& {Zhu}}]{2012RAA....12..735D}
{Deng}, L.-C., {Newberg}, H.~J., {Liu}, C., {et~al.} 2012, Research in
  Astronomy and Astrophysics, 12, 735, \dodoi{10.1088/1674-4527/12/7/003}

\bibitem[{{Dotter}(2016)}]{2016ApJS..222....8D}
{Dotter}, A. 2016, \apjs, 222, 8, \dodoi{10.3847/0067-0049/222/1/8}

\bibitem[{{Duch{\^e}ne} \& {Kraus}(2013)}]{2013ARA&A..51..269D}
{Duch{\^e}ne}, G., \& {Kraus}, A. 2013, \araa, 51, 269,
  \dodoi{10.1146/annurev-astro-081710-102602}

\bibitem[{{Duquennoy} \& {Mayor}(1991)}]{1991A&A...248..485D}
{Duquennoy}, A., \& {Mayor}, M. 1991, \aap, 500, 337

\bibitem[{{Eggleton}(2006)}]{2006epbm.book.....E}
{Eggleton}, P. 2006, {Evolutionary Processes in Binary and Multiple Stars}

\bibitem[{{El-Badry} {et~al.}(2018{\natexlab{a}}){El-Badry}, {Rix}, {Ting},
  {Weisz}, {Bergemann}, {Cargile}, {Conroy}, \& {Eilers}}]{2018MNRAS.473.5043E}
{El-Badry}, K., {Rix}, H.-W., {Ting}, Y.-S., {et~al.} 2018{\natexlab{a}},
  \mnras, 473, 5043, \dodoi{10.1093/mnras/stx2758}

\bibitem[{{El-Badry} {et~al.}(2018{\natexlab{b}}){El-Badry}, {Ting}, {Rix},
  {Quataert}, {Weisz}, {Cargile}, {Conroy}, {Hogg}, {Bergemann}, \&
  {Liu}}]{2018MNRAS.476..528E}
{El-Badry}, K., {Ting}, Y.-S., {Rix}, H.-W., {et~al.} 2018{\natexlab{b}},
  \mnras, 476, 528, \dodoi{10.1093/mnras/sty240}

\bibitem[{{Fu} {et~al.}(2020){Fu}, {Cat}, {Zong}, {Frasca}, {Gray}, {Ren},
  {Molenda-{\.Z}akowicz}, {Corbally}, {Catanzaro}, {Shi}, {Luo}, \&
  {Zhang}}]{2020RAA....20..167F}
{Fu}, J.-N., {Cat}, P.~D., {Zong}, W., {et~al.} 2020, Research in Astronomy and
  Astrophysics, 20, 167, \dodoi{10.1088/1674-4527/20/10/167}

\bibitem[{{Gaia Collaboration} {et~al.}(2016){Gaia Collaboration}, {Prusti},
  {de Bruijne}, {Brown}, {Vallenari}, {Babusiaux}, {Bailer-Jones}, {Bastian},
  {Biermann}, {Evans}, \& et~al.}]{2016A&A...595A...1G}
{Gaia Collaboration}, {Prusti}, T., {de Bruijne}, J.~H.~J., {et~al.} 2016,
  \aap, 595, A1, \dodoi{10.1051/0004-6361/201629272}

\bibitem[{{Gaia Collaboration} {et~al.}(2021){Gaia Collaboration}, {Brown},
  {Vallenari}, {Prusti}, {de Bruijne}, {Babusiaux}, {Biermann}, {Creevey},
  {Evans}, {Eyer}, \& et~al.}]{2021A&A...649A...1G}
{Gaia Collaboration}, {Brown}, A.~G.~A., {Vallenari}, A., {et~al.} 2021, \aap,
  649, A1, \dodoi{10.1051/0004-6361/202039657}

\bibitem[{{Gao} {et~al.}(2014){Gao}, {Liu}, {Zhang}, {Justham}, {Deng}, \&
  {Yang}}]{2014ApJ...788L..37G}
{Gao}, S., {Liu}, C., {Zhang}, X., {et~al.} 2014, \apjl, 788, L37,
  \dodoi{10.1088/2041-8205/788/2/L37}

\bibitem[{{Gao} {et~al.}(2017){Gao}, {Zhao}, {Yang}, \&
  {Gao}}]{2017MNRAS.469L..68G}
{Gao}, S., {Zhao}, H., {Yang}, H., \& {Gao}, R. 2017, \mnras, 469, L68,
  \dodoi{10.1093/mnrasl/slx048}

\bibitem[{{Green}(2018)}]{2018JOSS....3..695M}
{Green}, G. 2018, The Journal of Open Source Software, 3, 695,
  \dodoi{10.21105/joss.00695}

\bibitem[{{Green} {et~al.}(2019){Green}, {Schlafly}, {Zucker}, {Speagle}, \&
  {Finkbeiner}}]{2019ApJ...887...93G}
{Green}, G.~M., {Schlafly}, E., {Zucker}, C., {Speagle}, J.~S., \&
  {Finkbeiner}, D. 2019, \apj, 887, 93, \dodoi{10.3847/1538-4357/ab5362}

\bibitem[{{Guiglion} {et~al.}(2020){Guiglion}, {Matijevi{\v{c}}}, {Queiroz},
  {Valentini}, {Steinmetz}, {Chiappini}, {Grebel}, {McMillan}, {Kordopatis},
  {Kunder}, {Zwitter}, {Khalatyan}, {Anders}, {Enke}, {Minchev}, {Monari},
  {Wyse}, {Bienaym{\'e}}, {Bland-Hawthorn}, {Gibson}, {Navarro}, {Parker},
  {Reid}, {Seabroke}, \& {Siebert}}]{2020A&A...644A.168G}
{Guiglion}, G., {Matijevi{\v{c}}}, G., {Queiroz}, A.~B.~A., {et~al.} 2020,
  \aap, 644, A168, \dodoi{10.1051/0004-6361/202038271}

\bibitem[{{Halbwachs} {et~al.}(2003){Halbwachs}, {Mayor}, {Udry}, \&
  {Arenou}}]{2003A&A...397..159H}
{Halbwachs}, J.~L., {Mayor}, M., {Udry}, S., \& {Arenou}, F. 2003, \aap, 397,
  159, \dodoi{10.1051/0004-6361:20021507}

\bibitem[{{Han} {et~al.}(2020){Han}, {Ge}, {Chen}, \&
  {Chen}}]{2020RAA....20..161H}
{Han}, Z.-W., {Ge}, H.-W., {Chen}, X.-F., \& {Chen}, H.-L. 2020, Research in
  Astronomy and Astrophysics, 20, 161, \dodoi{10.1088/1674-4527/20/10/161}

\bibitem[{{Heintz}(1969)}]{1969JRASC..63..275H}
{Heintz}, W.~D. 1969, \jrasc, 63, 275

\bibitem[{{Hilditch}(2001)}]{2001icbs.book.....H}
{Hilditch}, R.~W. 2001, {An Introduction to Close Binary Stars}

\bibitem[{{Howell} {et~al.}(2014){Howell}, {Sobeck}, {Haas}, {Still},
  {Barclay}, {Mullally}, {Troeltzsch}, {Aigrain}, {Bryson}, {Caldwell},
  {Chaplin}, {Cochran}, {Huber}, {Marcy}, {Miglio}, {Najita}, {Smith},
  {Twicken}, \& {Fortney}}]{2014PASP..126..398H}
{Howell}, S.~B., {Sobeck}, C., {Haas}, M., {et~al.} 2014, \pasp, 126, 398,
  \dodoi{10.1086/676406}

\bibitem[{{Kirk} {et~al.}(2016){Kirk}, {Conroy}, {Pr{\v{s}}a}, {Abdul-Masih},
  {Kochoska}, {Matijevi{\v{c}}}, {Hambleton}, {Barclay}, {Bloemen}, {Boyajian},
  {Doyle}, {Fulton}, {Hoekstra}, {Jek}, {Kane}, {Kostov}, {Latham}, {Mazeh},
  {Orosz}, {Pepper}, {Quarles}, {Ragozzine}, {Shporer}, {Southworth},
  {Stassun}, {Thompson}, {Welsh}, {Agol}, {Derekas}, {Devor}, {Fischer},
  {Green}, {Gropp}, {Jacobs}, {Johnston}, {LaCourse}, {Saetre}, {Schwengeler},
  {Toczyski}, {Werner}, {Garrett}, {Gore}, {Martinez}, {Spitzer}, {Stevick},
  {Thomadis}, {Vrijmoet}, {Yenawine}, {Batalha}, \&
  {Borucki}}]{2016AJ....151...68K}
{Kirk}, B., {Conroy}, K., {Pr{\v{s}}a}, A., {et~al.} 2016, \aj, 151, 68,
  \dodoi{10.3847/0004-6256/151/3/68}

\bibitem[{{Kounkel} {et~al.}(2021){Kounkel}, {Covey}, {Stassun},
  {Price-Whelan}, {Holtzman}, {Chojnowski}, {Longa-Pe{\~n}a},
  {Rom{\'a}n-Z{\'u}{\~n}iga}, {Hernandez}, {Serna}, {Badenes}, {De Lee},
  {Majewski}, {Stringfellow}, {Kratter}, {Moe}, {Frinchaboy}, {Beaton},
  {Fern{\'a}ndez-Trincado}, {Mahadevan}, {Minniti}, {Beers}, {Schneider},
  {Barba}, {Brownstein}, {Garc{\'\i}a-Hern{\'a}ndez}, {Pan}, \&
  {Bizyaev}}]{2021AJ....162..184K}
{Kounkel}, M., {Covey}, K.~R., {Stassun}, K.~G., {et~al.} 2021, \aj, 162, 184,
  \dodoi{10.3847/1538-3881/ac1798}

\bibitem[{{LeCun} {et~al.}(1998){LeCun}, {Bottou}, {Bengio}, \&
  {Haffner}}]{1998PIEEELecun}
{LeCun}, Y., {Bottou}, L., {Bengio}, Y., \& {Haffner}, P. 1998, Proceedings of
  the IEEE, 86, 2278

\bibitem[{{Leung} \& {Bovy}(2019)}]{2019MNRAS.483.3255L}
{Leung}, H.~W., \& {Bovy}, J. 2019, \mnras, 483, 3255,
  \dodoi{10.1093/mnras/sty3217}

\bibitem[{{Li} {et~al.}(2021){Li}, {Shi}, {Yan}, {Fu}, {Li}, \&
  {Hou}}]{2021ApJS..256...31L}
{Li}, C.-q., {Shi}, J.-r., {Yan}, H.-l., {et~al.} 2021, \apjs, 256, 31,
  \dodoi{10.3847/1538-4365/ac22a8}

\bibitem[{{Li}(2020)}]{2020ApJ...892L..26L}
{Li}, G.-W. 2020, \apjl, 892, L26, \dodoi{10.3847/2041-8213/ab8123}

\bibitem[{{Lin} {et~al.}(2021){Lin}, {Wang}, {Mo}, {Xi}, {Zhang}, {Jiang},
  {Shi}, {Zhang}, {Zhang}, {Wei}, {Ye}, {Wu}, {Yan}, {Chen}, {Li}, {Li}, {Lin},
  {Lin}, {Sai}, {Xiang}, \& {Zhang}}]{2021MNRAS.tmp.2562L}
{Lin}, J., {Wang}, X., {Mo}, J., {et~al.} 2021, \mnras,
  \dodoi{10.1093/mnras/stab2812}

\bibitem[{{Liu}(2019)}]{2019MNRAS.490..550L}
{Liu}, C. 2019, \mnras, 490, 550, \dodoi{10.1093/mnras/stz2274}

\bibitem[{{Liu} {et~al.}(2020){Liu}, {Fu}, {Shi}, {Wu}, {Han}, {Chen}, {Dong},
  {Zhao}, {Chen}, {Zhang}, {Bai}, {Chen}, {Cui}, {Du}, {Hsia}, {Jiang}, {Hou},
  {Hou}, {Li}, {Li}, {Li}, {Liu}, {Liu}, {Luo}, {Ren}, {Tian}, {Tian}, {Wang},
  {Wu}, {Xie}, {Yan}, {Yang}, {Yu}, {Zhang}, {Zhang}, {Zhang}, {Zhang}, {Zhao},
  {Zhong}, {Zong}, \& {Zuo}}]{2020arXiv200507210L}
{Liu}, C., {Fu}, J., {Shi}, J., {et~al.} 2020, arXiv e-prints,
  arXiv:2005.07210.
\newblock \doarXiv{2005.07210}

\bibitem[{{Liu} {et~al.}(2019){Liu}, {Zhang}, {Howard}, {Bai}, {Lu}, {Soria},
  {Justham}, {Li}, {Zheng}, {Wang}, {Belczynski}, {Casares}, {Zhang}, {Yuan},
  {Dong}, {Lei}, {Isaacson}, {Wang}, {Bai}, {Shao}, {Gao}, {Wang}, {Niu},
  {Cui}, {Zheng}, {Mu}, {Zhang}, {Wang}, {Heger}, {Qi}, {Liao}, {Lattanzi},
  {Gu}, {Wang}, {Wu}, {Shao}, {Shen}, {Wang}, {Bregman}, {Di Stefano}, {Liu},
  {Han}, {Zhang}, {Wang}, {Ren}, {Zhang}, {Zhang}, {Wang}, {Cabrera-Lavers},
  {Corradi}, {Rebolo}, {Zhao}, {Zhao}, {Chu}, \& {Cui}}]{2019Natur.575..618L}
{Liu}, J., {Zhang}, H., {Howard}, A.~W., {et~al.} 2019, \nat, 575, 618,
  \dodoi{10.1038/s41586-019-1766-2}

\bibitem[{{Majewski} {et~al.}(2017){Majewski}, {Schiavon}, {Frinchaboy},
  {Allende Prieto}, {Barkhouser}, {Bizyaev}, {Blank}, {Brunner}, {Burton},
  {Carrera}, {Chojnowski}, {Cunha}, {Epstein}, {Fitzgerald}, {Garc{\'\i}a
  P{\'e}rez}, {Hearty}, {Henderson}, {Holtzman}, {Johnson}, {Lam}, {Lawler},
  {Maseman}, {M{\'e}sz{\'a}ros}, {Nelson}, {Nguyen}, {Nidever}, {Pinsonneault},
  {Shetrone}, {Smee}, {Smith}, {Stolberg}, {Skrutskie}, {Walker}, {Wilson},
  {Zasowski}, {Anders}, {Basu}, {Beland}, {Blanton}, {Bovy}, {Brownstein},
  {Carlberg}, {Chaplin}, {Chiappini}, {Eisenstein}, {Elsworth}, {Feuillet},
  {Fleming}, {Galbraith-Frew}, {Garc{\'\i}a}, {Garc{\'\i}a-Hern{\'a}ndez},
  {Gillespie}, {Girardi}, {Gunn}, {Hasselquist}, {Hayden}, {Hekker}, {Ivans},
  {Kinemuchi}, {Klaene}, {Mahadevan}, {Mathur}, {Mosser}, {Muna}, {Munn},
  {Nichol}, {O'Connell}, {Parejko}, {Robin}, {Rocha-Pinto}, {Schultheis},
  {Serenelli}, {Shane}, {Silva Aguirre}, {Sobeck}, {Thompson}, {Troup},
  {Weinberg}, \& {Zamora}}]{2017AJ....154...94M}
{Majewski}, S.~R., {Schiavon}, R.~P., {Frinchaboy}, P.~M., {et~al.} 2017, \aj,
  154, 94, \dodoi{10.3847/1538-3881/aa784d}

\bibitem[{{Matijevi{\v{c}}} {et~al.}(2010){Matijevi{\v{c}}}, {Zwitter},
  {Munari}, {Bienaym{\'e}}, {Binney}, {Bland-Hawthorn}, {Boeche}, {Campbell},
  {Freeman}, {Gibson}, {Gilmore}, {Grebel}, {Helmi}, {Navarro}, {Parker},
  {Seabroke}, {Siebert}, {Siviero}, {Steinmetz}, {Watson}, {Williams}, \&
  {Wyse}}]{2010AJ....140..184M}
{Matijevi{\v{c}}}, G., {Zwitter}, T., {Munari}, U., {et~al.} 2010, \aj, 140,
  184, \dodoi{10.1088/0004-6256/140/1/184}

\bibitem[{{Matijevi{\v{c}}} {et~al.}(2011){Matijevi{\v{c}}}, {Zwitter},
  {Bienaym{\'e}}, {Bland-Hawthorn}, {Freeman}, {Gilmore}, {Grebel}, {Helmi},
  {Munari}, {Navarro}, {Parker}, {Reid}, {Seabroke}, {Siebert}, {Siviero},
  {Steinmetz}, {Watson}, {Williams}, \& {Wyse}}]{2011AJ....141..200M}
{Matijevi{\v{c}}}, G., {Zwitter}, T., {Bienaym{\'e}}, O., {et~al.} 2011, \aj,
  141, 200, \dodoi{10.1088/0004-6256/141/6/200}

\bibitem[{{Mazzola} {et~al.}(2020){Mazzola}, {Badenes}, {Moe}, {Koposov},
  {Kounkel}, {Kratter}, {Covey}, {Walker}, {Thompson}, {Andrews}, {Freeman},
  {Anguiano}, {Carlberg}, {De Lee}, {Frinchaboy}, {Lewis}, {Majewski},
  {Nidever}, {Nitschelm}, {Price-Whelan}, {Roman-Lopes}, {Stassun}, \&
  {Troup}}]{2020MNRAS.499.1607M}
{Mazzola}, C.~N., {Badenes}, C., {Moe}, M., {et~al.} 2020, \mnras, 499, 1607,
  \dodoi{10.1093/mnras/staa2859}

\bibitem[{{Merle} {et~al.}(2017){Merle}, {Van Eck}, {Jorissen}, {Van der
  Swaelmen}, {Masseron}, {Zwitter}, {Hatzidimitriou}, {Klutsch}, {Pourbaix},
  {Blomme}, {Worley}, {Sacco}, {Lewis}, {Abia}, {Traven}, {Sordo}, {Bragaglia},
  {Smiljanic}, {Pancino}, {Damiani}, {Hourihane}, {Gilmore}, {Randich},
  {Koposov}, {Casey}, {Morbidelli}, {Franciosini}, {Magrini}, {Jofre},
  {Costado}, {Jeffries}, {Bergemann}, {Lanzafame}, {Bayo}, {Carraro},
  {Flaccomio}, {Monaco}, \& {Zaggia}}]{2017A&A...608A..95M}
{Merle}, T., {Van Eck}, S., {Jorissen}, A., {et~al.} 2017, \aap, 608, A95,
  \dodoi{10.1051/0004-6361/201730442}

\bibitem[{{Moe} \& {Di Stefano}(2017)}]{2017ApJS..230...15M}
{Moe}, M., \& {Di Stefano}, R. 2017, \apjs, 230, 15,
  \dodoi{10.3847/1538-4365/aa6fb6}

\bibitem[{{Moe} {et~al.}(2019){Moe}, {Kratter}, \&
  {Badenes}}]{2019ApJ...875...61M}
{Moe}, M., {Kratter}, K.~M., \& {Badenes}, C. 2019, \apj, 875, 61,
  \dodoi{10.3847/1538-4357/ab0d88}

\bibitem[{{Niu} {et~al.}(2021){Niu}, {Yuan}, {Wang}, \&
  {Liu}}]{2021arXiv210904031N}
{Niu}, Z., {Yuan}, H., {Wang}, S., \& {Liu}, J. 2021, arXiv e-prints,
  arXiv:2109.04031.
\newblock \doarXiv{2109.04031}

\bibitem[{{O'Briain} {et~al.}(2021){O'Briain}, {Ting}, {Fabbro}, {Yi}, {Venn},
  \& {Bialek}}]{2021ApJ...906..130O}
{O'Briain}, T., {Ting}, Y.-S., {Fabbro}, S., {et~al.} 2021, \apj, 906, 130,
  \dodoi{10.3847/1538-4357/abca96}

\bibitem[{{Pan} {et~al.}(2021){Pan}, {Fu}, {Zhang}, {Wang}, {Zong}, {Li}, \&
  {Zhang}}]{2021PASP..133d4202P}
{Pan}, Y., {Fu}, J.-N., {Zhang}, X., {et~al.} 2021, \pasp, 133, 044202,
  \dodoi{10.1088/1538-3873/abef77}

\bibitem[{{Pan} {et~al.}(2020){Pan}, {Fu}, {Zong}, {Zhang}, {Wang}, \&
  {Li}}]{2020ApJ...905...67P}
{Pan}, Y., {Fu}, J.-N., {Zong}, W., {et~al.} 2020, \apj, 905, 67,
  \dodoi{10.3847/1538-4357/abc250}

\bibitem[{{Pedregosa} {et~al.}(2012){Pedregosa}, {Varoquaux}, {Gramfort},
  {Michel}, {Thirion}, {Grisel}, {Blondel}, {M{\"u}ller}, {Nothman}, {Louppe},
  {Prettenhofer}, {Weiss}, {Dubourg}, {Vanderplas}, {Passos}, {Cournapeau},
  {Brucher}, {Perrot}, \& {Duchesnay}}]{2012arXiv1201.0490P}
{Pedregosa}, F., {Varoquaux}, G., {Gramfort}, A., {et~al.} 2012, arXiv
  e-prints, arXiv:1201.0490.
\newblock \doarXiv{1201.0490}

\bibitem[{{Pourbaix} {et~al.}(2004){Pourbaix}, {Tokovinin}, {Batten}, {Fekel},
  {Hartkopf}, {Levato}, {Morrell}, {Torres}, \& {Udry}}]{2004A&A...424..727P}
{Pourbaix}, D., {Tokovinin}, A.~A., {Batten}, A.~H., {et~al.} 2004, \aap, 424,
  727, \dodoi{10.1051/0004-6361:20041213}

\bibitem[{{Price-Whelan} {et~al.}(2017){Price-Whelan}, {Hogg},
  {Foreman-Mackey}, \& {Rix}}]{2017ApJ...837...20P}
{Price-Whelan}, A.~M., {Hogg}, D.~W., {Foreman-Mackey}, D., \& {Rix}, H.-W.
  2017, \apj, 837, 20, \dodoi{10.3847/1538-4357/aa5e50}

\bibitem[{{Price-Whelan} {et~al.}(2020){Price-Whelan}, {Hogg}, {Rix}, {Beaton},
  {Lewis}, {Nidever}, {Almeida}, {Badenes}, {Barba}, {Beers}, {Carlberg}, {De
  Lee}, {Fern{\'a}ndez-Trincado}, {Frinchaboy}, {Garc{\'\i}a-Hern{\'a}ndez},
  {Green}, {Hasselquist}, {Longa-Pe{\~n}a}, {Majewski}, {Nitschelm}, {Sobeck},
  {Stassun}, {Stringfellow}, \& {Troup}}]{2020ApJ...895....2P}
{Price-Whelan}, A.~M., {Hogg}, D.~W., {Rix}, H.-W., {et~al.} 2020, \apj, 895,
  2, \dodoi{10.3847/1538-4357/ab8acc}

\bibitem[{{Raghavan} {et~al.}(2010){Raghavan}, {McAlister}, {Henry}, {Latham},
  {Marcy}, {Mason}, {Gies}, {White}, \& {ten Brummelaar}}]{2010ApJS..190....1R}
{Raghavan}, D., {McAlister}, H.~A., {Henry}, T.~J., {et~al.} 2010, \apjs, 190,
  1, \dodoi{10.1088/0067-0049/190/1/1}

\bibitem[{{Ricker} {et~al.}(2015){Ricker}, {Winn}, {Vanderspek}, {Latham},
  {Bakos}, {Bean}, {Berta-Thompson}, {Brown}, {Buchhave}, {Butler}, {Butler},
  {Chaplin}, {Charbonneau}, {Christensen-Dalsgaard}, {Clampin}, {Deming},
  {Doty}, {De Lee}, {Dressing}, {Dunham}, {Endl}, {Fressin}, {Ge}, {Henning},
  {Holman}, {Howard}, {Ida}, {Jenkins}, {Jernigan}, {Johnson}, {Kaltenegger},
  {Kawai}, {Kjeldsen}, {Laughlin}, {Levine}, {Lin}, {Lissauer}, {MacQueen},
  {Marcy}, {McCullough}, {Morton}, {Narita}, {Paegert}, {Palle}, {Pepe},
  {Pepper}, {Quirrenbach}, {Rinehart}, {Sasselov}, {Sato}, {Seager},
  {Sozzetti}, {Stassun}, {Sullivan}, {Szentgyorgyi}, {Torres}, {Udry}, \&
  {Villasenor}}]{2015JATIS...1a4003R}
{Ricker}, G.~R., {Winn}, J.~N., {Vanderspek}, R., {et~al.} 2015, Journal of
  Astronomical Telescopes, Instruments, and Systems, 1, 014003,
  \dodoi{10.1117/1.JATIS.1.1.014003}

\bibitem[{{Salpeter}(1955)}]{1955ApJ...121..161S}
{Salpeter}, E.~E. 1955, \apj, 121, 161, \dodoi{10.1086/145971}

\bibitem[{{Schneider} {et~al.}(2001){Schneider}, {Ferrari}, {Matarrese}, \&
  {Portegies Zwart}}]{2001MNRAS.324..797S}
{Schneider}, R., {Ferrari}, V., {Matarrese}, S., \& {Portegies Zwart}, S.~F.
  2001, \mnras, 324, 797, \dodoi{10.1046/j.1365-8711.2001.04217.x}

\bibitem[{{Steinmetz} {et~al.}(2006){Steinmetz}, {Zwitter}, {Siebert},
  {Watson}, {Freeman}, {Munari}, {Campbell}, {Williams}, {Seabroke}, {Wyse},
  {Parker}, {Bienaym{\'e}}, {Roeser}, {Gibson}, {Gilmore}, {Grebel}, {Helmi},
  {Navarro}, {Burton}, {Cass}, {Dawe}, {Fiegert}, {Hartley}, {Russell},
  {Saunders}, {Enke}, {Bailin}, {Binney}, {Bland-Hawthorn}, {Boeche}, {Dehnen},
  {Eisenstein}, {Evans}, {Fiorucci}, {Fulbright}, {Gerhard}, {Jauregi}, {Kelz},
  {Mijovi{\'c}}, {Minchev}, {Parmentier}, {Pe{\~n}arrubia}, {Quillen}, {Read},
  {Ruchti}, {Scholz}, {Siviero}, {Smith}, {Sordo}, {Veltz}, {Vidrih}, {von
  Berlepsch}, {Boyle}, \& {Schilbach}}]{2006AJ....132.1645S}
{Steinmetz}, M., {Zwitter}, T., {Siebert}, A., {et~al.} 2006, \aj, 132, 1645,
  \dodoi{10.1086/506564}

\bibitem[{{Tian} {et~al.}(2018){Tian}, {Liu}, {Yuan}, {Chen}, {Xiang}, {Huang},
  {Wang}, {Zhang}, {Guo}, {Ren}, {Huo}, {Yang}, {Zhang}, {Bi}, {Yang}, {Liu},
  {Zhang}, {Li}, {Wu}, \& {Zhang}}]{2018RAA....18...52T}
{Tian}, Z.-J., {Liu}, X.-W., {Yuan}, H.-B., {et~al.} 2018, Research in
  Astronomy and Astrophysics, 18, 052, \dodoi{10.1088/1674-4527/18/5/52}

\bibitem[{{Ting} {et~al.}(2019){Ting}, {Conroy}, {Rix}, \&
  {Cargile}}]{2019ApJ...879...69T}
{Ting}, Y.-S., {Conroy}, C., {Rix}, H.-W., \& {Cargile}, P. 2019, \apj, 879,
  69, \dodoi{10.3847/1538-4357/ab2331}

\bibitem[{{Traven} {et~al.}(2017){Traven}, {Matijevi{\v{c}}}, {Zwitter},
  {{\v{Z}}erjal}, {Kos}, {Asplund}, {Bland-Hawthorn}, {Casey}, {De Silva},
  {Freeman}, {Lin}, {Martell}, {Schlesinger}, {Sharma}, {Simpson}, {Zucker},
  {Anguiano}, {Da Costa}, {Duong}, {Horner}, {Hyde}, {Kafle}, {Munari},
  {Nataf}, {Navin}, {Reid}, \& {Ting}}]{2017ApJS..228...24T}
{Traven}, G., {Matijevi{\v{c}}}, G., {Zwitter}, T., {et~al.} 2017, \apjs, 228,
  24, \dodoi{10.3847/1538-4365/228/2/24}

\bibitem[{{Traven} {et~al.}(2020){Traven}, {Feltzing}, {Merle}, {Van der
  Swaelmen}, {{\v{C}}otar}, {Church}, {Zwitter}, {Ting}, {Sahlholdt},
  {Asplund}, {Bland-Hawthorn}, {De Silva}, {Freeman}, {Martell}, {Sharma},
  {Zucker}, {Buder}, {Casey}, {D'Orazi}, {Kos}, {Lewis}, {Lin}, {Lind},
  {Simpson}, {Stello}, {Munari}, \& {Wittenmyer}}]{2020A&A...638A.145T}
{Traven}, G., {Feltzing}, S., {Merle}, T., {et~al.} 2020, \aap, 638, A145,
  \dodoi{10.1051/0004-6361/202037484}

\bibitem[{{von Hippel} {et~al.}(1994){von Hippel}, {Storrie-Lombardi},
  {Storrie-Lombardi}, \& {Irwin}}]{1994MNRAS.269...97V}
{von Hippel}, T., {Storrie-Lombardi}, L.~J., {Storrie-Lombardi}, M.~C., \&
  {Irwin}, M.~J. 1994, \mnras, 269, 97, \dodoi{10.1093/mnras/269.1.97}

\bibitem[{{Wang} {et~al.}(2021{\natexlab{a}}){Wang}, {Fu}, {Niu}, {Pan}, {Li},
  {Zong}, \& {Hou}}]{2021MNRAS.504.4302W}
{Wang}, J., {Fu}, J., {Niu}, H., {et~al.} 2021{\natexlab{a}}, \mnras, 504,
  4302, \dodoi{10.1093/mnras/stab1219}

\bibitem[{{Wang} {et~al.}(2021{\natexlab{b}}){Wang}, {Fu}, {Zong}, {Wang}, \&
  {Zhang}}]{2021MNRAS.506.6117W}
{Wang}, J., {Fu}, J.-N., {Zong}, W., {Wang}, J., \& {Zhang}, B.
  2021{\natexlab{b}}, \mnras, 506, 6117, \dodoi{10.1093/mnras/stab1705}

\bibitem[{{Wang} {et~al.}(2020){Wang}, {Luo}, {Chen}, {Hou}, {Zhang}, {Zhao},
  {Li}, {Hou}, \& {LAMOST MRS Collaboration}}]{2020ApJ...891...23W}
{Wang}, R., {Luo}, A.~L., {Chen}, J.-J., {et~al.} 2020, \apj, 891, 23,
  \dodoi{10.3847/1538-4357/ab6dea}

\bibitem[{{Wang} {et~al.}(2021{\natexlab{c}}){Wang}, {Zhang}, {Bai}, {Yuan},
  {Xiang}, {Zhang}, {Hou}, {Zuo}, {Du}, {Li}, {Yang}, {Cui}, {Wang}, {Li},
  {Kovalev}, {Li}, {Tian}, {Zong}, {Han}, {Liu}, {Luo}, {Shi}, {Fu}, {Bi},
  {Han}, \& {Liu}}]{2021arXiv210903149W}
{Wang}, S., {Zhang}, H., {Bai}, Z., {et~al.} 2021{\natexlab{c}}, arXiv
  e-prints, arXiv:2109.03149.
\newblock \doarXiv{2109.03149}

\bibitem[{{Xiang} {et~al.}(2019){Xiang}, {Ting}, {Rix}, {Sandford}, {Buder},
  {Lind}, {Liu}, {Shi}, \& {Zhang}}]{2019ApJS..245...34X}
{Xiang}, M., {Ting}, Y.-S., {Rix}, H.-W., {et~al.} 2019, \apjs, 245, 34,
  \dodoi{10.3847/1538-4365/ab5364}

\bibitem[{{Yang} {et~al.}(2020){Yang}, {Long}, {Shan}, {Zhang}, {Guo}, {Bai},
  {Bai}, {Cui}, {Wang}, \& {Liu}}]{2020ApJS..249...31Y}
{Yang}, F., {Long}, R.~J., {Shan}, S.-S., {et~al.} 2020, \apjs, 249, 31,
  \dodoi{10.3847/1538-4365/ab9b77}

\bibitem[{{Yang} {et~al.}(2021{\natexlab{a}}){Yang}, {Zhang}, {Long}, {Lu},
  {Shan}, {Wei}, {Fu}, {Zhang}, {Zhao}, {Bai}, {Yi}, {Zheng}, {Zhou}, \&
  {Liu}}]{2021arXiv211007944Y}
{Yang}, F., {Zhang}, B., {Long}, R.~J., {et~al.} 2021{\natexlab{a}}, arXiv
  e-prints, arXiv:2110.07944.
\newblock \doarXiv{2110.07944}

\bibitem[{{Yang} {et~al.}(2021{\natexlab{b}}){Yang}, {Wu}, {Luo}, \&
  {Zou}}]{2021FrASS...8...59Y}
{Yang}, W.-Y., {Wu}, K.-F., {Luo}, A.~L., \& {Zou}, Z.-Q. 2021{\natexlab{b}},
  Frontiers in Astronomy and Space Sciences, 8, 59,
  \dodoi{10.3389/fspas.2021.634328}

\bibitem[{{Yuan} {et~al.}(2015){Yuan}, {Liu}, {Xiang}, {Huang}, {Chen}, {Wu},
  {Hou}, \& {Zhang}}]{2015ApJ...799..135Y}
{Yuan}, H., {Liu}, X., {Xiang}, M., {et~al.} 2015, \apj, 799, 135,
  \dodoi{10.1088/0004-637X/799/2/135}

\bibitem[{{Zhang}(2020{\natexlab{a}})}]{2020zndo...4381163Z}
{Zhang}, B. 2020{\natexlab{a}}, {hypergravity/berliner: A toolkit for stellar
  tracks and isochrones.}, 2020.1221.0,  Zenodo, \dodoi{10.5281/zenodo.4381163}

\bibitem[{{Zhang}(2020{\natexlab{b}})}]{2020zndo...4381155Z}
---. 2020{\natexlab{b}}, {hypergravity/laspec: A toolkit for LAMOST spectra.},
  2020.1221.0,  Zenodo, \dodoi{10.5281/zenodo.4381155}

\bibitem[{{Zhang} {et~al.}(2020{\natexlab{a}}){Zhang}, {Liu}, \&
  {Deng}}]{2020ApJS..246....9Z}
{Zhang}, B., {Liu}, C., \& {Deng}, L.-C. 2020{\natexlab{a}}, \apjs, 246, 9,
  \dodoi{10.3847/1538-4365/ab55ef}

\bibitem[{{Zhang} {et~al.}(2020{\natexlab{b}}){Zhang}, {Liu}, {Li}, {Deng},
  {Yan}, \& {Shi}}]{2020RAA....20...51Z}
{Zhang}, B., {Liu}, C., {Li}, C.-Q., {et~al.} 2020{\natexlab{b}}, Research in
  Astronomy and Astrophysics, 20, 051, \dodoi{10.1088/1674-4527/20/4/51}

\bibitem[{{Zhang} {et~al.}(2021){Zhang}, {Li}, {Yang}, {Xiong}, {Fu}, {Liu},
  {Tian}, {Li}, {Wang}, {Liang}, {Zhou}, {Zong}, {Yang}, {Liu}, \&
  {Hou}}]{2021ApJS..256...14Z}
{Zhang}, B., {Li}, J., {Yang}, F., {et~al.} 2021, \apjs, 256, 14,
  \dodoi{10.3847/1538-4365/ac0834}

\bibitem[{{Zhang} {et~al.}(2015){Zhang}, {Ge}, {Lu}, {Cao}, {Chen}, {Mao}, \&
  {Jiang}}]{2015PASP..127.1292Z}
{Zhang}, J.-C., {Ge}, L., {Lu}, X.-M., {et~al.} 2015, \pasp, 127, 1292,
  \dodoi{10.1086/684369}

\bibitem[{{Zong} {et~al.}(2020){Zong}, {Fu}, {De Cat}, {Wang}, {Shi}, {Luo},
  {Zhang}, {Frasca}, {Molenda-{\.Z}akowicz}, {Gray}, {Corbally}, {Catanzaro},
  {Cang}, {Wang}, {Chen}, {Hou}, {Liu}, {Niu}, {Pan}, {Tian}, {Yan}, {Zhang},
  \& {Zuo}}]{2020ApJS..251...15Z}
{Zong}, W., {Fu}, J.-N., {De Cat}, P., {et~al.} 2020, \apjs, 251, 15,
  \dodoi{10.3847/1538-4365/abbb2d}

\end{thebibliography}
\bibliographystyle{aasjournal}

\end{document}